\documentclass[aps,onecolumn,prd,amssymb,eqsecnum,nofootinbib,superscriptaddress,showpacs,floatfix,amsmath]{revtex4}
\newlength{\picwidth}
\usepackage{epsfig}
\usepackage{amsmath}
\usepackage{dcolumn}
\usepackage{amssymb}
\usepackage{amsfonts}
\usepackage{bm}
\def\be{\begin{equation}}
\def\ee{\end{equation}}
\def\ba{\begin{eqnarray}}
\def\ea{\end{eqnarray}}

\def\di{\Delta_i}
\def\hfit{H_{fit}}
\def\omm{\Omega_M}
\def\epsm{\epsilon_M}
\def\oml{\Omega_\Lambda}
\def\epsl{\epsilon_\Lambda}
\def\sigi{\sigma^2_i}
\def\chit{{\tilde\chi^2}}
\def\ffav{\langle F_i^2\rangle}
\def\dfav{\langle \di F_i\rangle}
\def\dffav{\langle\di F_i^2\rangle}
\def\dfffav{\langle\di F_i^3\rangle}
\def\gfav{\langle G_iF_i\rangle}
\def\ifav{\langle I_iF_i\rangle}
\def\dgav{\langle\di G_i\rangle}
\def\ggav{\langle G_i^2\rangle}
\def\igav{\langle I_iG_i\rangle}
\def\diav{\langle \di I_i\rangle}
\def\iiav{\langle I_i^2\rangle}

\begin{document}

\title{Systematic corrections to the measured cosmological constant as a result of local inhomogeneity}

\author{R. Ali Vanderveld}
\affiliation{Center for Radiophysics and Space Research, Cornell University, Ithaca, NY 14853}
\author{\'{E}anna \'{E}. Flanagan}
\affiliation{Center for Radiophysics and Space Research, Cornell University, Ithaca, NY 14853}
\affiliation{Laboratory for Elementary Particle Physics, Cornell University, Ithaca, NY 14853}
\author{Ira Wasserman}
\affiliation{Center for Radiophysics and Space Research, Cornell University, Ithaca, NY 14853}
\affiliation{Laboratory for Elementary Particle Physics, Cornell University, Ithaca, NY 14853}
\date{\today}
\begin{abstract}

We explicitly calculate the lowest order systematic inhomogeneity-induced corrections to the cosmological constant that one would infer from an analysis of the luminosities and redshifts of Type Ia supernovae, assuming a homogeneous universe. The calculation entails a post-Newtonian expansion within the framework of second order perturbation theory, wherein we consider the effects of subhorizon density perturbations in a flat, dust dominated universe.  Within this formalism, we calculate luminosity distances and redshifts along the past light cone of an observer.  The luminosity distance-redshift relation is then averaged over viewing angles and ensemble averaged, assuming that density fluctuations at a given cosmic time are a homogeneous random process.  The resulting relation is fit to that of a homogeneous model containing dust and a cosmological constant, in order to deduce the best-fit cosmological constant density $\Omega_{\Lambda}$.  We find that the luminosity distance-redshift relation is indeed modified, even for large sample sizes, but only by a very small fraction, of order $10^{-5}$ for $z\sim 0.1$. This lowest order deviation depends on the peculiar velocities of the source and the observer. However, when fitting this perturbed relation to that of a homogeneous universe, via maximizing a likelihood function, we find that the inferred cosmological constant can be surprisingly large, depending on the range of redshifts sampled.  For a sample of supernovae extending from $z_{min}=0.02$ out to a limiting redshift $z_{max}=0.15$, we find that $\Omega_{\Lambda}\approx 0.004$.  The value of $\Omega_{\Lambda}$ has a large variance, and its magnitude tends to get progressively larger as the limiting redshift $z_{max}$ gets smaller, implying that precision measurements of $\Omega_{\Lambda}$ from nearby supernova data will require taking this effect into account. This effect has been referred to in the past as the ``fitting problem", and more recently as subhorizon ``backreaction". We find that it is likely too small to explain the observed value $\Omega_{\Lambda}\approx 0.7$.  There have been previous claims of much larger backreaction effects.  By contrast to those calculations, our work is directly related to how observers deduce cosmological parameters from astronomical data.

\end{abstract}
\pacs{ 98.80.-k, 98.80.Jk, 98.80.Es }
\maketitle

\section{Introduction}

It appears as though the Universe is expanding at an accelerating rate, as has been deduced from luminosity distance measurements of Type Ia supernovae, which appear dimmer than one would expect based on general relativity without a cosmological constant \cite{Riess, Perlmutter}.  This acceleration has also been deduced from measurements of the current matter density $\Omega_{M}\approx 0.27$, which is too small to close the Universe as required by cosmic microwave background radiation (CMB) observations with $H_0$ priors from HST \cite{Bennett, web}.  Explanations for this discrepancy have been put forward, but most employ a modification of general relativity on cosmological scales or the addition of an exotic ``dark energy" field.

There have also been attempts to explain this seemingly anomalous cosmic acceleration as a consequence of subhorizon inhomogeneity, rather than modified gravity or dark energy.  A perturbation is referred to as ``subhorizon" if its wavelength is small compared to the Hubble length: $\lambda/L_H \ll 1$.  It has been suggested that small-scale density perturbations could cause the appearance of accelerated expansion without the need to introduce any form of dark energy, which is an appealing prospect \cite{Rasanen, Notari, Kolb1, Kolb2}.  The fact that inhomogeneity can systematically modify our interpretation of cosmological measurements was first realized by Ellis, who called it the ``fitting problem" \cite{Ellis, Ellis2}.  The basic idea is this: Due to the nonlinearity of the Einstein equation, the operators for taking spatial averages and for time evolution do not commute.  This means that, although our universe is homogeneous in the mean, it will likely not have the same time evolution as that of the corresponding homogeneous universe.  Nevertheless, we routinely fit distance data to FRW models, a procedure that introduces errors into the inferred properties of our Universe, and these errors will be present even for very large samples of Type Ia supernovae.  

Our goal in this paper is to calculate the lowest order fitting effect by calculating the cosmological constant density $\Omega_{\Lambda}$ that one would deduce from a perturbed luminosity distance-redshift relation $D_L(z)$.  If we treat cosmological fluctuations perturbatively and as a random process as suggested by the ``fair sample hypothesis" \cite{Peebles2}, then this fitting effect should be fundamentally nonlinear in the density contrast $\delta=(\rho-\langle\rho\rangle)/\langle\rho\rangle$, requiring that we work to at least second order in $\delta$. This is because the ensemble averages of first order quantities vanish.  We model observations out to some moderate redshift $z_{max}\sim 0.1 \ll 1$.  Within the corresponding comoving spherical region, the Hubble flow velocity $v_H$ is bounded above by
\be
\frac{v_H}{c}\lesssim z_{max} \sim 0.1~,
\ee
allowing us to use post-Newtonian expansions.  There are two different velocity scales that occur, the Hubble flow velocity $v_H$ and the peculiar velocity $v_p$.  The corresponding dimensionless small parameters are
\be
\varepsilon_H=\frac{v_H}{c}\sim\frac{H_0 r}{c}\lesssim z_{max}\sim 0.1
\ee
and
\be
\varepsilon_p=\frac{v_p}{c}\sim\delta\left(\frac{H_0\lambda_c}{c}\right)~,
\ee
where $\lambda_c\sim 10~{\rm Mpc}$ is the wavelength of the dominant perturbation mode.  In our computation, we will treat both of these parameters as being of formally the same order, and we will denote both by ``$\varepsilon$" for book keeping purposes.  At the end of our computation we can identify terms that scale as $\varepsilon_H^n\varepsilon_p^m$ for different values of $m$ and $n$.  As mentioned above, we also expand separately in the fractional density perturbation $\delta$.  We will compute redshifts $z(\lambda)$ and luminosity distances $H_0D_L(\lambda)$ as functions of the affine parameter $\lambda$ to third order in $\varepsilon$ and to second order in $\delta$.  Combining these results to eliminate $\lambda$ will yield $D_L$ as a function of $z$.

\begin{figure}
\begin{center}
\epsfig{file=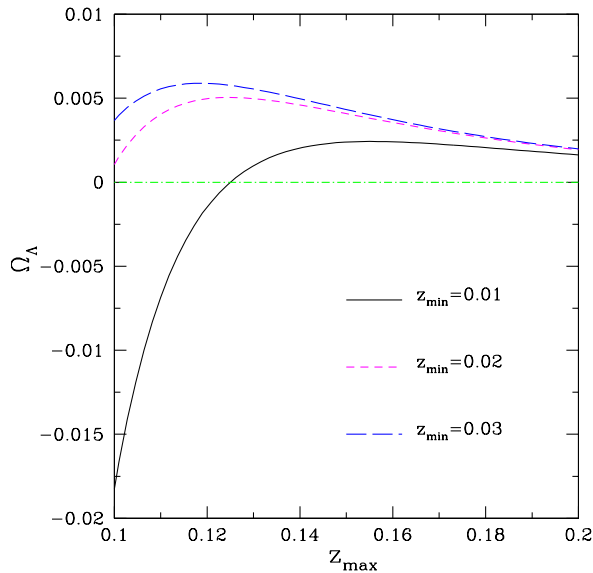,width=\picwidth,height =8cm}
\caption{The best-fit cosmological constant density $\Omega_{\Lambda}$ plotted as a function of the maximum redshift $z_{max}$, for the choices $z_{min}=0.01,~0.02,$ and $0.03$.  The horizontal dash-dot line shows the actual model value $\Omega_{\Lambda}=0$.}  
\label{omega}
\end{center}
\end{figure}

Using this expansion method, we find that the lowest order inhomogeneity-induced correction to the luminosity distance scales as $|\Delta D_L|/D_L\sim \delta^2(H_0\lambda_c/c) \sim 10^{-5}$.  We then fit this relation to what one would expect from a homogeneous cosmological model which contains dust with a density $\Omega_{M}$ and a cosmological constant with a density $\Omega_{\Lambda}$,
\be
D_L(z)=\frac{1+z}{H_0\sqrt{|1-\Omega_{M}-\Omega_{\Lambda}|}}{\cal F}\left[\sqrt{|1-\Omega_{M}-\Omega_{\Lambda}|}\int^z_0\frac{dz'}{\sqrt{\Omega_{M}(1+z')^3+\left(1-\Omega_M-\Omega_{\Lambda}\right)(1+z')^2+\Omega_{\Lambda}}}\right]~,
\ee
by maximizing a likelihood function.  Here ${\cal F}(u)=u$ for a flat universe, ${\cal F}(u)=\sinh(u)$ for an open universe, and ${\cal F}(u)=\sin(u)$ for a closed universe.  We find that the result for the cosmological constant density is dependent on the size of the redshift range for which we have supernova data.  These results are summarized in Figure~\ref{omega}.  
For data from $z_{min}=0.02$ out to a limiting redshift $z_{max}=0.15$, we find that the best-fit cosmological constant density is $\Omega_{\Lambda}\approx 0.004$, and $\Omega_{\Lambda}$ tends to get larger as $z_{max}$ gets smaller.  The best-fit $|\Omega_{\Lambda}|$ also becomes larger as $z_{min}$ becomes smaller, since $|\Delta D_L|/D_L$ becomes large on small scales.  Although this ensemble averaged result is still quite small, we find that the variance can be $\sigma_{\Lambda}^2\sim 1$ for a sample of 100 supernovae out to a redshift $z_{max}\sim 0.2$.  One implication of these results is that precision measurements of the cosmological constant from nearby supernova data require that we measure $D_L(z)$ over a large enough redshift range, with a large enough sample.  One could also try to correct for some of the effects of inhomogeneity, using available information about large scale structure and about our own peculiar velocity \cite{Velocities, Velocities2}.

The analysis presented here is more realistic than similar analyses within the context of simplified models of structure formation, such as the spherically symmetric Lema\^{i}tre-Tolman-Bondi (LTB) models \cite{Bondi, Celerier, INN, Garfinkle}, Swiss cheese models \cite{Kantowski} and their variants \cite{Biswas, Kai, Brouzakis1, Brouzakis2}.  This is because we look at the full three dimensional problem, and assume that there are no bulk flows on cosmological length scales.  There have also been analyses of the perturbations to the luminosity distance-redshift relation that go to Newtonian order \cite{Frieman, HW, Wang, Durrer}, that only consider superhorizon perturbation modes \cite{Flanagan, Hirata}, and that use Taylor expansions of the luminosity distance \cite{BMR}, which are most appropriate for long-wavelength perturbations.  In contrast, we go to post-Newtonian order, we only consider subhorizon modes, and we fit to FRW models, so that we may fully address the ``fitting problem".  

Our analysis is also fundamentally different from those in Refs. \cite{Rasanen, Notari, Kolb1, Kolb2, Buchert1, Buchert2}, as we choose a different method for obtaining averaged expansion parameters. These authors average the expansion rate over a constant time slice, whereas we choose to calculate only observable quantities, namely the luminosity distance and the redshift, along the past light cone of the observer.  We then combine these expressions into $D_L(z,\theta,\phi)$, average over viewing angles and ensemble average, and then fit the results to what one would expect in a homogeneous model containing dust and a cosmological constant to find the best-fit value for $\Omega_{\Lambda}$.  This approach better simulates the process of gathering and analyzing supernova data, and it leads to a different result with a stable perturbative expansion.

Refs. \cite{Rasanen, Notari, Kolb1, Kolb2} base their characterization of the expansion rate of the Universe on quantities that are not related to how observers have deduced the existence of dark energy.  In these papers, perturbations are spatially averaged over a constant time slice.  Such a spatial average is somewhat arbitrary, as it is dependent on the choice of spatial hypersurface.  This is in contrast to the observable significance of $D_L(z)$.  Refs. \cite{Notari, Kolb1, Kolb2} also use the synchronous gauge for their calculations, wherein there are metric perturbations of order $\delta$.  Since $\delta\gtrsim 1$ on small scales, this gauge is particularly ill suited to perturbation theory.  In contrast, in Newtonian-type gauges the metric perturbation is of order $\delta(H_0\lambda_c/c)^2\ll\delta$.  We explore this difference in Section VI.

The organization of the paper is as follows: In Section II below, we introduce our coordinate choice, wherein we recast the Friedmann-Robertson-Walker (FRW) metric as an expansion around flat space, and in Section III we present the fundamental post-Newtonian optics equations that we will need for this calculation.  We then explain our method of computation and calculate the necessary unperturbed quantities in Section IV.  Here we also compute the luminosity distances and redshifts for a perturbed matter dominated universe, finding $z$ and $H_0D_L$ to second order in $\delta$ and to third order in $\varepsilon$, and we find that we may write the lowest order correction to $D_L(z)$ in terms of the peculiar velocity field.  We then fit to a homogeneous model in Section V to find the best-fit $\Omega_{\Lambda}$ and its variance.  The detailed redshift and luminosity distance equations are in Appendix A, the necessary results of second order perturbation theory are reviewed in Appendix B, and the averaging is discussed in Appendix C.  Then, in Section VI we discuss the previous results in the synchronous gauge and show that one can choose coordinates and a definition of ``acceleration" such that it appears as though there could be a larger fitting effect. We argue that such a result would be unphysical.  A detailed discussion of transforming to synchronous coordinates is given in Appendix D.  Finally, in Section VII we make our concluding remarks.  As usual, Greek indices will be summed over all four spacetime dimensions while Latin indices will be summed only over the three spatial dimensions.  We will also write 3-vectors in boldface and put arrows over 4-vectors.

\section{Post-Newtonian expansion of the local FRW metric}

In general, certain coordinate choices allow us to conveniently recast the local metric as an expansion around flat space, as was first emphasized for the FRW metric by Peebles \cite{Peebles}.  We will take advantage of such an expansion so that we may use the standard post-Newtonian formalism for this calculation.  Starting with the usual FRW metric with $c=G=1$,
\be
ds^2=-d\tau^2+a^2(\tau)\left(d\chi^2+\chi^2d\Omega^2\right)~,
\ee
we can define the new radial coordinate
\be
\tilde{r}=a(\tau)\chi
\ee
so that the line element becomes
\be
ds^2=-\left(1-H^2\tilde{r}^2\right)d\tau^2-2H\tilde{r}d\tau d\tilde{r}+d\tilde{r}^2+\tilde{r}^2d\Omega^2~,
\ee
where the Hubble parameter of a flat and dust-dominated FRW universe is $H(\tau)=(1/a)(da/d\tau)=2/3\tau$; we will specialize to this case for the remainder of this paper.  Now we change coordinates to the standard post-Newtonian gauge.  In this gauge, the metric to first post-Newtonian order can be written as
\be
ds^2=g_{\mu\nu}dx^{\mu}dx^{\nu}=-\left(1+2\Phi+2\Phi^2\right)dt^2+2\zeta_idx^idt+\left(1-2\Phi\right)\gamma_{ij}dx^idx^j~,
\label{basicmetric}
\ee
where $\gamma_{ij}$ is a flat spatial metric, the potential $\Phi$ contains both Newtonian and post-Newtonian pieces, $\zeta_i$ is the usual gravitomagnetic potential, and
\be
3\dot{\Phi}+{\bf \nabla}\cdot{\bf \zeta}=0
\ee 
is the gauge condition.  Achieving this form for the metric entails transforming from $\tau$ and $\tilde{r}$ to $t$ and $r$, defined by
\be
\tau=t\left[1-\frac{r^2}{3t^2}-\frac{r^4}{30t^4}+O\left(\frac{r^6}{t^6}\right)\right]
\ee
and
\be
\tilde{r}=r\left[1-\frac{r^2}{9t^2}+O\left(\frac{r^4}{t^4}\right)\right]~.
\ee
Then the line element becomes
\be
ds^2=-\left[1+\frac{2r^2}{9t^2}+\frac{46r^4}{405t^4}+O\left(\frac{r^6}{t^6}\right)\right]dt^2+\left[\frac{4r^3}{15t^3}+O\left(\frac{r^5}{t^5}\right)\right]drdt+\left[1-\frac{2r^2}{9t^2}+O\left(\frac{r^4}{t^4}\right)\right]\left(dr^2+r^2d\Omega^2\right)
\label{metricbackground}
\ee
to the necessary order in $r$.  This metric is of the post-Newtonian form (\ref{basicmetric}) if we define
\be
\Phi_{(0)}=\frac{r^2}{9t^2}+\frac{2r^4}{45t^4}
\ee
and 
\be
\zeta_{r(0)}=\frac{2r^3}{15t^3}~.
\ee  
Here subscripts ``(0)" denote unperturbed, background quantities; we will add cosmological perturbations in subsequent sections. The unperturbed density in the new coordinates is
\be
\rho_{(0)}=\frac{1}{6\pi t^2}\left[1+\frac{2r^2}{3t^2}+O\left(\frac{r^4}{t^4}\right)\right]~,
\ee
and the continuity equation tells us that the unperturbed 3-velocity must be of the form ${\bf v}_{(0)}=v_{(0)}\partial/\partial r$, where
\be
v_{(0)}=\frac{2r}{3t}\left[1+\frac{r^2}{9t^2}+O\left(\frac{r^4}{t^4}\right)\right]~,
\ee
and where $v_{(0)}=dr/dt$. Thus, we see that counting orders of $\varepsilon\sim v/c$ is equivalent to counting orders of $r/t$ in these coordinates. Our coordinate choice and expansion method also have the consequence that the analysis of this paper is only valid for small redshifts.  

In general in the standard post-Newtonian gauge, the connection coefficients are
\be
\Gamma^t_{tt}=\dot{\Phi}~,
\label{gamma1}
\ee
\be
\Gamma^t_{ti}=\Phi_{,i}~,
\ee
\be
\Gamma^t_{ij}=-\dot{\Phi}\gamma_{ij}-\zeta_{(i|j)}~,
\ee
\be
\Gamma^i_{tt}=\gamma^{ij}\Phi_{,j}~,
\ee
\be
\Gamma^i_{tj}=-\dot{\Phi}\delta^i_{~j}+\gamma^{ik}\zeta_{[k|j]}~,
\ee
and
\be
\Gamma^i_{jk}=\tilde{\Gamma}^i_{jk}-\Phi_{,k}\delta^i_{~j}-\Phi_{,j}\delta^i_{~k}+\Phi_{,l}\gamma^{il}\gamma_{jk}~,
\label{gammalast}
\ee
to the necessary order in $\varepsilon$, where $\tilde{\Gamma}^i_{jk}$ is the connection associated with the flat spatial metric $\gamma_{ij}$, which we will choose to be that of standard spherical coordinates $(r,\theta,\phi)$, as in Ref. \cite{MTW}.  Vertical bars represent covariant derivatives with respect to $\gamma_{ij}$.  We will also need the Ricci tensor components
\be
R_{tt}=\nabla^2\Phi
\ee
and
\be
R_{ij}=\nabla^2\Phi\delta_{ij}~.
\ee

Furthermore, the first post-Newtonian hydrodynamic and Einstein equations are
\be
\frac{\partial}{\partial t}\left[\rho\left(1+\frac{v^2}{2}-3\Phi\right)\right]+{\bf \nabla}\cdot\left[\rho\left(1+\frac{v^2}{2}-3\Phi\right){\bf v}\right]=0~,
\ee
\be
\frac{\partial{\bf v}}{\partial t}+\left({\bf v}\cdot{\bf \nabla}\right){\bf v}=-{\bf \nabla}\left(\Phi+2\Phi^2\right)-\dot{{\bf \zeta}}-\left({\bf \nabla}\times{\bf \zeta}\right)\times{\bf v}+3\dot{\Phi}{\bf v}+4{\bf v}\left({\bf v}\cdot{\bf \nabla}\right)\Phi-v^2{\bf \nabla}\Phi~,
\ee
\be
\nabla^2\Phi=4\pi\rho\left(1+2v^2-2\Phi\right)~,
\ee
and
\be
\nabla^2{\bf \zeta}=16\pi\rho{\bf v}+{\bf \nabla}\dot{\Phi}~,
\ee
in this gauge.  The 3-velocity ${\bf v}$ is related to the 4-velocity $\vec{u}$ of the fluid by
\be
\vec{u}=\left(u^t,u^i\right)\equiv\gamma\left(1,v^i\right)~,
\ee
where demanding that $\vec{u}\cdot\vec{u}=-1$ yields
\be
\gamma^2=1+v^2-2\Phi+2\Phi^2-6\Phi v^2+v^4+2{\bf \zeta}\cdot{\bf v}~.
\label{gammafactor}
\ee

\section{Computation of luminosity distance and redshift}

\subsection{Computing $D_L(z)$ in a general spacetime}

In this section we will review how to compute luminosity distances and redshifts in a general spacetime, as in Refs. \cite{Sachs, MTW}.  Our analysis is initially similar to that of Ref. \cite{BMR}, although they eventually rely on Taylor expansions around the observer's location.  Such expansions are sensible for long-wavelength perturbations, but not for the short-wavelength perturbations that we consider here.  We focus attention on a particular observer at some event $\cal{P}$.  In our application to perturbed FRW spacetimes, this observer will be at $r=0$ and at $t=t_0$ for some fixed $t_0$.  We consider the congruence of geodesics forming this observer's past light cone.  Given the connection, we then find ray trajectories $x^{\alpha}(\lambda)$ by noting that the 4-momentum is $k^{\alpha}=dx^{\alpha}/d\lambda$, and by using the geodesic equation
\be
\frac{dk^{\alpha}}{d\lambda}=k^{\beta}\partial_{\beta}k^{\alpha}=-\Gamma^{\alpha}_{\mu\nu}k^{\mu}k^{\nu}~,
\label{nullgeo}
\ee
where we have defined $d/d\lambda=k^{\alpha}\partial_{\alpha}$. Here the affine parameter $\lambda$ is chosen such that $\lambda=0$ at the observer and $\lambda=\lambda_s<0$ at the source.  We also note that the 4-momentum is null. 

The expansion $\theta$ of the congruence of null rays is related to the area ${\cal A}(\lambda)$ of a bundle of rays by
\be
\theta=\frac{1}{{\cal A}}\frac{d{\cal A}\left(\lambda\right)}{d\lambda}~.
\ee
We can find $\theta$ by using the Raychadhuri equation
\be
\frac{d\theta}{d\lambda}=-R_{\mu\nu}k^{\mu}k^{\nu}-\frac{1}{2}\theta^{2}-2|\sigma|^2~,
\label{raychaud}
\ee
where we have defined the shear of the congruence
\be
|\sigma|^2=\frac{1}{2}\left[k_{\alpha;\beta}k^{\alpha;\beta}-\frac{1}{2}\theta^2\right]~,
\ee
and where we require $\theta\sim 2/\lambda$ as $\lambda\rightarrow 0$, so that the area of the beam goes to zero at $\lambda=0$.  The shear $\sigma\equiv\sqrt{|\sigma|^2}$ is given by the differential equation
\be
\frac{d\sigma}{d\lambda}=-\sigma\theta+C_{\alpha\beta\mu\nu}k^{\alpha}k^{\nu}\bar{t}^{\mu}\bar{t}^{\beta}~,
\label{shear}
\ee
where $C_{\alpha\beta\mu\nu}$ is the Weyl tensor, and we have defined a null Newman-Penrose tetrad composed of the real 4-vectors $k^{\mu}$ and $m^{\mu}$, and the complex conjugate 4-vectors $t^{\mu}$ and $\bar{t}^{\mu}$.  These satisfy the orthogonality conditions
\be
k^{\mu}m_{\mu}=\bar{t}^{\mu}t_{\mu}=1
\ee
and
\be
k^{\mu}k_{\mu}=m^{\mu}m_{\mu}=t^{\mu}t_{\mu}=k^{\mu}t_{\mu}=m^{\mu}t_{\mu}=0~,
\ee
as in \cite{BMR}.  They are chosen at the observer and then extended along each geodesic in the congruence by parallel transport.  We also choose the initial condition $\sigma=0$ at $\lambda=0$.  

Once we find $\theta$, we then find the luminosity distance as a function of the affine parameter at the source,
\ba
D_{L}(\lambda_s)&=&\lim_{\Delta\lambda\rightarrow 0}\left[-\Delta\lambda\left(1+z\right)^2\exp\left(\frac{1}{2}\int^{\lambda_s}_{\Delta\lambda}\theta d\lambda\right)\right]\nonumber\\
&=&-\lambda_s\left(1+z\right)^2\exp\left[\frac{1}{2}\int^{\lambda_s}_0\left(\theta-\frac{2}{\lambda}\right)d\lambda\right]
\label{DLinitial}
\ea
where $\Delta\lambda$ corresponds to the size of the observer's telescope, which we set to zero.  The right hand side of Eq. (\ref{DLinitial}) has a well defined, finite, limit as $\Delta\lambda\rightarrow 0$ due to the aforementioned initial condition placed on $\theta$.  Note also that the right hand side has an overall minus sign due to our convention that the affine parameter is negative.

The redshift observed at $\lambda=0$, of the light emitted from the source at $\lambda_s$, is
\be
1+z(\lambda_s)=\frac{\left(u_{\alpha}k^{\alpha}\right)_s}{\left(u_{\beta}k^{\beta}\right)_o}~,
\label{Zinitial}
\ee
where
\be
u_{\alpha}k^{\alpha}=\gamma\left(g_{tt}k^t+g_{ti}k^i+g_{it}v^ik^t+g_{ij}v^ik^j\right)~,
\ee
and where the subscript ``s" will in general denote quantities evaluated at the source at the emission time and the subscript ``o" will denote quantities evaluated at the observer at the observation time.  By combining Eqs. (\ref{DLinitial}) and (\ref{Zinitial}) we can, in principle, compute $D_L$ as a function of $z$ in a general spacetime.

\subsection{Computing $D_L(z)$ to first post-Newtonian order}

Now we specialize the results of the preceding subsection to a perturbed FRW metric in the post-Newtonian gauge (\ref{basicmetric}).  Our goal is to find both $H_0D_L$ and $z$ to order $\varepsilon^3$.  At the observer, we have chosen $r=0$ and $t=t_0$ and we have normalized the 4-momentum such that $k^r=-1$.  This implies that $\lambda\approx -r$ and $r/t\sim -\lambda/t\sim\varepsilon$ to lowest order.  We will thus need to find the right hand side of Eq. (\ref{DLinitial}) to order $\lambda\varepsilon^2$ so that we may find $H_0D_L$ to order $\varepsilon^3$.  Because of this, we see that we will need the integral in the exponential to order $\varepsilon^2$, and therefore we will need to find $\lambda\theta$ to order $\varepsilon^2$.  Similarly, inspection of Eq. (\ref{Zinitial}) tells us to what post-Newtonian order we will need to compute the components of $k^{\alpha}$.  To lowest order, $g_{tt}\sim 1$, $g_{ti}=g_{it}\sim\varepsilon^3$, $g_{ij}\sim 1$, $\gamma\sim 1$, and $v^i\sim\varepsilon$, and therefore we will need $k^t$ to order $\varepsilon^3$ and we will need the spatial components $k^i$ to order $\varepsilon^2$.

The post-Newtonian pieces of $k^{\alpha}$ must be as small or smaller than order $\varepsilon^2$, as can be seen by noting that $\lambda\Gamma^{\alpha}_{\mu\nu}\sim\varepsilon^2$ in the null geodesic equation (\ref{nullgeo}).  Given this assumption and the normalization of $k^{\alpha}$, Eq. (\ref{nullgeo}) reduces to
\be
\frac{dk^{\alpha}}{d\lambda}=\frac{\partial k^{\alpha}}{\partial t}-\frac{\partial k^{\alpha}}{\partial r}+O\left(\frac{\varepsilon^4}{\lambda}\right)=-\Gamma^{\alpha}_{tt}+2\Gamma^{\alpha}_{tr}-\Gamma^{\alpha}_{rr}+O\left(\frac{\varepsilon^4}{\lambda}\right)~.
\ee
Plugging in the connection coefficients from Eq. (\ref{gamma1})-(\ref{gammalast}), we find
\be
\frac{dk^t}{d\lambda}=2\Phi_{,r}+\zeta_{r,r}+O\left(\frac{\varepsilon^4}{\lambda}\right)~,
\ee
\be
\frac{dk^r}{d\lambda}=O\left(\frac{\varepsilon^3}{\lambda}\right)~,
\ee
\be
\frac{d}{d\lambda}\left(r k^{\theta}\right)=-\frac{2}{r}\Phi_{,\theta}+O\left(\frac{\varepsilon^3}{\lambda}\right)~,
\ee
and
\be
\frac{d}{d\lambda}\left(r k^{\phi}\right)=-\frac{2}{r\sin^2\theta}\Phi_{,\phi}+O\left(\frac{\varepsilon^3}{\lambda}\right)~.
\ee
Using the specified initial conditions, the solutions to these equations are
\be
k^t=1-2\Phi-\zeta_{r}-2\int_0^r\dot{\Phi}dr'+O\left(\varepsilon^4\right)~,
\label{kt}
\ee
\be
k^r=-1+O\left(\varepsilon^3\right)~,
\label{kr}
\ee
\be
k^{\theta}=\frac{2}{r}\int_0^r\frac{dr'}{r'}\Phi_{,\theta}+O\left(\varepsilon^3\right)~,
\label{ktheta}
\ee
and
\be
k^{\phi}=\frac{2}{r\sin^2\theta}\int_0^r\frac{dr'}{r'}\Phi_{,\phi}+O\left(\varepsilon^3\right)~;
\label{kphi}
\ee
the integrals above are performed along the unperturbed ray, where $t(\lambda)=t_0+\lambda$ and $r(\lambda)=-\lambda$.  We can then find the perturbed ray trajectory by integrating Eqs. (\ref{kt})-(\ref{kphi}) with respect to $\lambda$.  Most notably, Eq. (\ref{kr}) leads to $\lambda=-r+O(\lambda\varepsilon^3)$.  This means that we can easily rewrite Eq. (\ref{DLinitial}) in terms of the radial coordinate $r$ of the source:
\be
D_L=r\left(1+z\right)^2\exp\left[-\frac{1}{2}\int^{r}_0\left(\theta+\frac{2}{r'}\right)dr'\right]+O\left(r\varepsilon^3\right)~.
\label{DLagain}
\ee

In order to find the expansion $\theta$, we first need to find the shear, given by Eq. (\ref{shear}).  The solution to this equation is
\be
\sigma=\frac{1}{\lambda^2}\int_0^{\lambda}\left(\lambda'\right)^2C_{\alpha\beta\mu\nu}k^{\alpha}k^{\nu}\bar{t}^{\mu}\bar{t}^{\beta}d\lambda'~;
\ee
since $|k|\sim|\bar{t}|\sim 1$, it turns out that the lowest order shear is $\sigma\sim \varepsilon^2/\lambda$.  Inserting $|\sigma|^2\sim\varepsilon^4/\lambda^2$ into the Raychaudhuri equation (\ref{raychaud}) gives a contribution of order $\varepsilon^4/\lambda$ to the expansion $\theta$.  However, we already know that we only need $\theta$ to order $\varepsilon^2/\lambda$, and so this contribution is negligible for our purposes here.  Neglecting shear and defining $\delta\theta=\theta-2/\lambda$, we rewrite Eq. (\ref{raychaud}) as
\ba
\frac{d(\delta\theta)}{d\lambda}&=&-R_{tt}-R_{rr}-\frac{2}{\lambda}(\delta\theta)+O\left(\frac{\varepsilon^3}{\lambda^2}\right)\nonumber\\
&=&-2\nabla^2\Phi-\frac{2}{\lambda}(\delta\theta)+O\left(\frac{\varepsilon^3}{\lambda^2}\right)~.
\ea
The solution to this is
\be
\delta\theta=\frac{2}{r^2}\int^{r}_{0}\left(r'\right)^2\nabla^2\Phi dr'+O\left(\frac{\varepsilon^3}{\lambda}\right)~,
\label{deltatheta}
\ee
where we are using $\lambda=-r+O(\lambda\varepsilon^3)$.  Using this result in Eq. (\ref{DLagain}) yields our final result for the post-Newtonian luminosity distance
\be
D_L=r\left(1+z\right)^2\left[1-\int^r_0\frac{dr'}{r'^2}\int^{r'}_0\left(r''\right)^2\nabla^2\Phi dr''\right]+O\left(r\varepsilon^3\right)~.
\label{DLend}
\ee

We now turn to evaluating the redshift $z$ as a function of the affine parameter $\lambda$.  Equation (\ref{Zinitial}) is the general expression for the redshift, and it depends on $u_{\alpha}k^{\alpha}$ at the source and at the observer.  To order $\varepsilon^3$, using Eqs. (\ref{basicmetric}), (\ref{gammafactor}), and our solutions for $k^{\alpha}$, we obtain
\ba
u_{\alpha}k^{\alpha}&=&g_{\alpha\beta}u^{\alpha}k^{\beta}\nonumber\\
&=&-1-v^r-\frac{1}{2}v^2+\Phi+3v^r\Phi-\frac{1}{2}v^rv^2+2\int_0^r\dot{\Phi}dr'+v_{\theta}k^{\theta}+v_{\phi}k^{\phi}+O\left(\varepsilon^4\right)~,
\ea
where $k^{\theta}$ and $k^{\phi}$ are given by Eqs. (\ref{ktheta}) and (\ref{kphi}), respectively.  Therefore, the post-Newtonian redshift is
\ba
1+z&=&\frac{\left(u_{\alpha}k^{\alpha}\right)_s}{\left(u_{\beta}k^{\beta}\right)_o}\nonumber\\
&=&1+v_s^r-v_o^r+\Phi_o-\Phi_s+\frac{1}{2}\left(v_s^2-v_o^2\right)+\left(v_o^r\right)^2-v_o^rv_s^r-2\int^r_0\dot{\Phi}dr'+(v_{\theta}k^{\theta}+v_{\phi}k^{\phi})_o-(v_{\theta}k^{\theta}+v_{\phi}k^{\phi})_s\nonumber\\
&+&\Phi_ov_o^r+\Phi_sv_o^r+\Phi_ov_s^r-3\Phi_sv_s^r-\frac{1}{2}v_o^2\left(v_s^r-v_o^r\right)+\left(v_o^r\right)^2\left(v_s^r-v_o^r\right)+\frac{1}{2}v_s^2\left(v_s^r-v_o^r\right)+O\left(\varepsilon^4\right)~.
\label{Z}
\ea
In Eqs. (\ref{DLend}) and (\ref{Z}), the right hand sides are evaluated at $r=-\lambda$ and $t=t_0+\lambda$.  Recall that subscripts ``o" denote quantities evaluated at the observer where $r=0$ and $t=t_0$, while subscripts ``s" denote quantities evaluated at the source $(t(\lambda),r(\lambda),\theta,\phi)$.

\section{Adding density perturbations}

\subsection{Basic method}

In this section we apply the formalism of Section III to a spherical region in a perturbed FRW spacetime.  We will describe that region using the post-Newtonian metric (\ref{basicmetric}).  We expand the metric functions $\Phi$ and $\zeta^i$ and the fluid 3-velocity $v^i$ in powers of the density contrast $\delta$ as 
\be
\Phi=\Phi_{(0)}+\Phi_{(1)}+\Phi_{(2)}+O\left(\delta^3\right)~,
\ee
\be
\zeta_i=\zeta_{i(0)}+\zeta_{i(1)}+\zeta_{i(2)}+O\left(\delta^3\right)~,
\ee
and
\be
v^i=v^i_{(0)}+v^i_{(1)}+v^i_{(2)}+O\left(\delta^3\right)~,
\ee
respectively. We also expand the null geodesic $x^{\alpha}$ and 4-momentum $k^{\alpha}=dx^{\alpha}/d\lambda$ as
\be
x^{\alpha}=x_{(0)}^{\alpha}+x_{(1)}^{\alpha}+x_{(2)}^{\alpha}+O\left(\delta^3\right)
\ee
and
\be
k^{\alpha}=k_{(0)}^{\alpha}+k_{(1)}^{\alpha}+k_{(2)}^{\alpha}+O\left(\delta^3\right)~,
\ee
respectively.  For the remainder of the paper, quantities that are zeroth order in $\delta$ will be denoted by a subscript ``(0)", first order by a subscript ``(1)", and second order by a subscript ``(2)".  Also henceforth ``first order" and ``second order" will always refer to orders in $\delta$, not $\varepsilon$, unless otherwise specified.

In the perturbed spacetime, we will calculate the redshift $z$ and luminosity distance $D_L$ as functions of the observation time $t_0$, of the affine parameter $\lambda$ along the past-directed null geodesic, and of the 4-momentum $\vec{k}$ of photons at $r=0$ and $t=t_0$.  We parameterize this future-directed null vector $\vec{k}$ in terms of angles $\theta$ and $\phi$, in such a way that $k^r=-1$ and $\vec{k}$ is in the direction $(\theta,\phi)$ at $r=0$. We can thus express $D_L$ and $z$ as functions of $\lambda$, $\theta$, and $\phi$ at fixed $t_0$, and by eliminating the affine parameter $\lambda$ we can compute $D_L(z,\theta,\phi)$. 

We can then take an average over angles to find $D_L(z)$, where we must take some care since there are two sets of relevant angles.  There are the angles $(\tilde{\theta},\tilde{\phi})$ which parameterize the direction of $\vec{k}$ in the observer's rest frame, and then there are the coordinate angles $(\theta,\phi)$.  We will need to average over $(\tilde{\theta},\tilde{\phi})$.  This means that we will need to know the relationship between the related infinitesimal solid angles $d\Omega^2$ and $d\tilde{\Omega}^2$.  We define Cartesian coordinates $(x^1,x^2,x^3)$ in terms of the polar coordinates $(r,\theta,\phi)$ in the standard way.  An orthonormal set of basis vectors for the observer's local Lorentz frame can be obtained by renormalizing the coordinate basis vectors $\partial/\partial t$ and $\partial/\partial x^i$ and boosting.  The result is
\be
\vec{e}_t=\left[1+\frac{1}{2}v_o^2-\Phi_o+O\left(\varepsilon^3\right)\right]\frac{\partial}{\partial t}+\left[v_o^i+O\left(\varepsilon^3\right)\right]\frac{\partial}{\partial x^i}
\label{basist}
\ee
and
\be
\vec{e}_i=\left[v_o^i+O\left(\varepsilon^3\right)\right]\frac{\partial}{\partial t}+\left[\delta_{ij}\left(1+\Phi_o\right)+\frac{1}{2}v_o^iv_o^j++O\left(\varepsilon^3\right)\right]\frac{\partial}{\partial x^j}~.
\label{basisi}
\ee
The angles $(\theta,\phi)$ are defined by
\be
\vec{k}=k^t\frac{\partial}{\partial t}-n^i\frac{\partial}{\partial x^i}~,
\label{coordk}
\ee
with
\be
{\bf n}=\left(\sin\theta\cos\phi, \sin\theta\sin\phi, \cos\theta\right)~,
\ee
while the observer's angles $(\tilde{\theta},\tilde{\phi})$ are defined by
\be
\vec{k}\propto \vec{e}_t-\tilde{n}^i\vec{e}_i~,
\label{orthok}
\ee
with
\be
{\bf \tilde{n}}=\left(\sin\tilde{\theta}\cos\tilde{\phi}, \sin\tilde{\theta}\sin\tilde{\phi}, \cos\tilde{\theta}\right)~.
\ee
By inserting (\ref{basist}) and (\ref{basisi}) into (\ref{orthok}) and then comparing with (\ref{coordk}), we find
\be
{\bf n}\propto {\bf\tilde{n}}+\Phi_o{\bf\tilde{n}}-{\bf v}_o+\frac{1}{2}\left({\bf v}_o\cdot{\bf\tilde{n}}\right){\bf v}_o+O\left(\varepsilon^3\right)~.
\ee
This gives
\be
d^2\tilde{\Omega}=d^2\Omega\left[1-2\left({\bf v}_o\cdot{\bf n}\right)+O\left(\varepsilon^2\right)\right]~.
\label{Jacobian}
\ee

After averaging over viewing angles, we find the expected value of $D_L(z)$ by taking an ensemble average, wherein we treat the density perturbation $\delta$ at any fixed time as a homogeneous random process.  Once we have the averaged $D_L(z)$, we can then analyze these data in terms of a homogeneous universe to see if we would find an apparent acceleration.  Assuming Gaussian uncertainties, we perform a chi-squared fit to a FRW model with a matter density $\Omega_{M}$ and a cosmological constant density $\Omega_{\Lambda}$.

\subsection{Unperturbed quantities}

In the unperturbed background, everything is spherically symmetric, and the line element in our coordinates is given by Eq. (\ref{metricbackground}). The background four-momentum $k_0^{\alpha}$ is purely in the $t-r$ plane, and is given by Eqs. (\ref{kt}) and (\ref{kr}) to be
\be
k_{(0)}^t(r,t)=1-\frac{2r^2}{9t^2}+\frac{2r^3}{135t^3}+O\left(\frac{r^4}{t^4}\right)
\ee
and
\be
k_{(0)}^r(r,t)=-1-\frac{4r^3}{27t^3}+O\left(\frac{r^4}{t^4}\right)~.
\ee
Since $k_{(0)}^t=dt/d \lambda$ and $k_{(0)}^r=dr/d \lambda$, we can integrate and invert these equations to find the unperturbed ray trajectory; keeping in mind the conditions that $r=\lambda=0$ and $t=t_0$ at the observer, we find
\be
t(\lambda)=t_0+\lambda\left[1-\frac{2\lambda^2}{27t_0^2}+O\left(\frac{\lambda^3}{t_0^3}\right)\right]
\ee
and
\be
r(\lambda)=-\lambda\left[1+O\left(\frac{\lambda^3}{t_0^3}\right)\right]
\ee
in the unperturbed background.

Using this, we can use the solution (\ref{deltatheta}) to the Raychaudhuri equation to find the background expansion $\theta_{(0)}$,
\be
\theta_{(0)}(\lambda)=\frac{2}{\lambda}-\frac{4}{9t_0^2}\lambda+O\left(\frac{\lambda^2}{t_0^3}\right)~.
\ee
Then the zeroth-order luminosity distance is given by Eq. (\ref{DLend}) to be
\ba
D_{L(0)}&=&\left(1+z\right)^2r\left(1-\frac{r^2}{9t^2}\right)+O\left(\frac{r^4}{t^3}\right)\nonumber\\
&=&\frac{2}{3H_0}\left(1+z\right)^2\frac{r}{t}\left[1-\frac{r}{t}+\frac{8r^2}{9t^2}+O\left(\frac{r^3}{t^3}\right)\right]~,
\label{dlzero}
\ea
where we have defined $H_0=2/3t_0$.  The zeroth-order redshift is found from Eq. (\ref{Z})~,
\be
z_{(0)}=\frac{2r}{3t}+\frac{r^2}{9t^2}+\frac{4r^3}{27t^3}+O\left(\frac{r^4}{t^4}\right)~,
\label{zzero}
\ee
and we eventually find the expected $D_{L(0)}(z)$ by inverting Eq. (\ref{zzero}) and plugging the result into Eq. (\ref{dlzero}):
\be
D_{L(0)}(z)=\frac{z}{H_0}\left[1+\frac{1}{4}z-\frac{1}{8}z^2+O\left(z^3\right)\right]~.
\label{dlzero2}
\ee
Thus, for the background, the best-fit cosmological constant density is $\Omega_{\Lambda}=0$ and the deceleration parameter is $q_0=1/2$.

\subsection{Second order perturbed optics}

The perturbed post-1-Newtonian line element is, from Eq. (\ref{basicmetric}),
\ba
ds^2\approx&-&\left(1+2\Phi_{(0)}+2\Phi_{(0)}^2+2\Phi_{(1)}+4\Phi_{(0)}\Phi_{(1)}+2\Phi_{(2)}+2\Phi_{(1)}^2+4\Phi_{(0)}\Phi_{(2)}\right)dt^2
\nonumber\\
&+&2\left(\zeta_{i(0)}+\zeta_{i(1)}+\zeta_{i(2)}\right)dx^idt+\left(1-2\Phi_{(0)}-2\Phi_{(1)}-2\Phi_{(2)}\right)\gamma_{ij}dx^idx^j
\ea
and the perturbed luminosity distance (\ref{DLend}) is defined to be
\be
D_L=(1+z)^2 E_L~,
\ee
where
\ba
E_{L}(r,\theta,\phi)&=&E_{L(0)}(r,\theta,\phi)+E_{L(1)}(r,\theta,\phi)+E_{L(2)}(r,\theta,\phi)+O\left(\delta^3\right)\nonumber\\
&=&r\left[1-\int^r_0\frac{dr'}{r'^2}\int^{r'}_0\left(r''\right)^2\nabla^2\left(\Phi_{(0)}+\Phi_{(1)}+\Phi_{(2)}\right) dr''\right]+O\left(\delta^3\right)~,
\ea
and where we have pulled out the factor of $(1+z)^2$ for simplicity.  We then find that the order $\delta$ perturbation is
\be
E_{L(1)}=-\frac{2}{3H_0}\frac{r}{t_0}\int^r_0\frac{dr'}{r'^2}\int^{r'}_0\left(r''\right)^2\nabla^2\Phi_{(1)} dr''~,
\label{dlfirst}
\ee
and the order $\delta^2$ perturbation is
\be
E_{L(2)}=-\frac{2}{3H_0}\frac{r}{t_0}\int^r_0\frac{dr'}{r'^2}\int^{r'}_0\left(r''\right)^2\nabla^2\Phi_{(2)} dr''~,
\label{dlsecond}
\ee
where $H_0=2/3t_0$. In general, all of the terms involving potentials and velocities in these equations, and in those that follow, are evaluated along the zeroth-order, unperturbed, geodesic.

We can now calculate the perturbed redshift 
\be
z(r,\theta,\phi)=z_{(0)}(r,\theta,\phi)+z_{(1)}(r,\theta,\phi)+z_{(2)}(r,\theta,\phi)+O(\delta^3)
\ee 
from Eq. (\ref{Z}), using our knowledge of the zeroth-order quantities, to find
\ba
z_{(1)}&=&v^r_{s(1)}-v^r_{o(1)}+\Phi_{o(1)}-\Phi_{s(1)}+\frac{2r}{3t}\left(v^r_{s(1)}-v^r_{o(1)}\right)-2\int^r_0\dot{\Phi}_{(1)}dr'+\frac{2r}{3t}\Phi_{o(1)}\nonumber\\
&-&\frac{2r}{t}\Phi_{s(1)}-\frac{r^2}{9t^2}v^r_{o(1)}+\frac{r^2}{3t^2}v^r_{s(1)}+O\left(\delta\varepsilon^4\right)
\label{zfirst}
\ea
and
\ba
z_{(2)}&=&v^r_{s(2)}-v^r_{o(2)}+\Phi_{o(2)}-\Phi_{s(2)}+\frac{2r}{3t}\left(v^r_{s(2)}-v^r_{o(2)}\right)+\frac{1}{2}\left(v_{s(1)}^2-v_{o(1)}^2\right)+\left(v^r_{o(1)}\right)^2\nonumber\\
&-&v^r_{o(1)}v^r_{s(1)}-2\int^r_0\dot{\Phi}_{(2)}dr'+\left(v_{\theta(1)}k_{(1)}^{\theta}+v_{\phi(1)}k_{(1)}^{\phi}\right)_o-\left(v_{\theta(1)}k_{(1)}^{\theta}+v_{\phi(1)}k_{(1)}^{\phi}\right)_s+\frac{2r}{3t}\Phi_{o(2)}\nonumber\\
&-&\frac{2r}{t}\Phi_{s(2)}-\frac{r^2}{9t^2}v^r_{o(2)}+\frac{r^2}{3t^2}v^r_{s(2)}+\frac{r}{3t}\left[v_{s(1)}^2-v_{o(1)}^2\right]+\frac{2r}{3t}\left[\left(v^r_{s(1)}\right)^2+\left(v^r_{o(1)}\right)^2-v^r_{s(1)}v^r_{o(1)}\right]\nonumber\\
&+&\Phi_{o(1)}v^r_{o(1)}+\Phi_{s(1)}v^r_{o(1)}+\Phi_{o(1)}v^r_{s(1)}-3\Phi_{s(1)}v^r_{s(1)}+x_{(1)}^iv^r_{s(1),i}+O\left(\delta^2\varepsilon^4\right)~,
\label{zsecond}
\ea
where the first order perturbation to the null geodesic is
\be
x_{(1)}^i=-\int_0^{r}k_{(1)}^i dr'~.  
\ee
All of the quantities above are evaluated along the zeroth-order geodesic, and the integrals are performed along an unperturbed central ray where $r(\lambda)=-\lambda$ and $t(\lambda)=t_0+\lambda$.

Now we have found the redshift $z$ and luminosity distance $H_0D_L$ as functions of affine parameter $\lambda$ and initial 4-momentum $\vec{k}_o$, to second order in $\delta$ and to third order in $\varepsilon$. Adding the redshift equations (\ref{zzero}), (\ref{zfirst}), and (\ref{zsecond}) yields $z(\lambda,\theta,\phi)$.  Similarly, the luminosity distance $D_L(\lambda,\theta,\phi)$ is found from adding Eqs. (\ref{dlzero2}), (\ref{dlfirst}), and (\ref{dlsecond}), after replacing the factors of $(1+z)^2$.  Inverting $z(\lambda,\theta,\phi)$ perturbatively, in terms of either $\delta$ or $\varepsilon$, gives us $\lambda$ as a function of $z$.  Plugging this into $D_L(\lambda,\theta,\phi)$ yields an expression for $D_L(z,\theta,\phi)$.  We then angle average this and then ensemble average, assuming that density fluctuations at a given cosmic time are a homogeneous random process.  Details of this full procedure are given in Appendices A, B, and C, and the result is
\be
D_L(z)=\frac{z}{H_0}\left(1+\frac{1}{4}z-\frac{1}{8}z^2\right)+\Delta D_L(z)~,
\label{finalanswer}
\ee
where $\Delta D_L(z)$ depends on the two point correlation function.  We will only need the lowest order piece of this, which is
\be
\Delta D_L(z)=-\frac{1}{3H_0^2}f'\left(\frac{z}{H_0}\right)\langle v^2_{o(1)} \rangle+O\left(\frac{f\varepsilon^2\delta^2}{H_0}\right)~,
\label{deltalow}
\ee
using $z\approx H_0r$ to lowest order.  The function $f$ is related to the velocity two point correlation function (see Appendix C):
\be
f(r)=\frac{3\left\langle {\bf n}\cdot{\bf v}({\bf r}_0,t){\bf n}\cdot{\bf v}({\bf r}_0+r{\bf n},t) \right\rangle}{\left\langle |{\bf v}({\bf r}_0,t)|^2\right\rangle}-1~,
\label{feqn}
\ee
where ${\bf n}$ is a unit vector that defines the viewing direction and ${\bf r}_0$ is an arbitrary location in space.  Note that $f(r)$ is independent of time, even though $\langle{\bf v}({\bf r}_0,t)^2\rangle$ does depend on time.  This is because the time dependences of the numerator and denominator cancel.

The perturbation to the luminosity distance is proportional to
\be
\langle v_{o(1)}^2 \rangle = \frac{4}{9H_0^2}\langle \left(\nabla\Phi_{o(1)} \right)^2 \rangle~;
\ee
this qualitative scaling has been argued for in Refs. \cite{HS1} and \cite{Futamase}.  We can Fourier transform $\Phi_{(1)}$, in terms of a wavevector $k^i$ (not to be confused with the previously-defined 4-momentum) \cite{Kolb1},
\be
\Phi_{(1)} = \int\frac{d^3k}{(2\pi)^3}\Phi_{{\bf k}}e^{i{\bf k}\cdot{\bf r}}
\label{ft}
\ee
so that we may write the average of $(\nabla\Phi_{(1)})^2$ as a sum over modes:
\be
\langle \left(\nabla\Phi_{(1)}\right)^2 \rangle = \frac{9}{4}H_0^4\int_0^{\infty}\frac{dk}{k^3}\Delta^2(k) ~,
\label{fourier}
\ee
where $\Delta(k)$ is the dimensionless power spectrum of matter density fluctuations at the present time, defined by
\be
\langle\delta^2\rangle=\int_{-\infty}^{\infty}d(\ln k)\Delta(k)^2~.
\label{deltaeqn}
\ee
We adopt the following power spectrum
\be
\Delta^2(k)=C^2\left(\frac{k}{H_0}\right)^4T^2\left(\frac{k}{k_{eq}}\right)~,
\label{HZ}
\ee
where the factor of $(k/H_0)^4$ reflects a Harrison-Zel'dovich flat spectrum, the amplitude $C=1.9\times 10^{-5}$ is set by observations, and $T(y)$ is the transfer function.  The BBKS transfer function \cite{BBKS} is a good fit for $T$ in the absence of dark energy,
\be
T(y)=\frac{\ln\left(1+2.34y\right)}{2.34y}\left[1+3.89y+\left(16.1y\right)^2+\left(5.46y\right)^3+\left(6.71y\right)^4\right]^{-1/4}~,
\label{transfer}
\ee
where 
\be
y=\frac{k}{k_{eq}}=\frac{k\theta^{1/2}}{\Omega_{X}{\rm h}^2{\rm Mpc}^{-1}}~.
\label{yeqn}
\ee
Here we show the most general form of the transfer function, where $\theta=\rho_{ER}/1.68\rho_{\gamma}$ (not to be confused with the expansion $\theta$) is the density of relativistic particles divided by the density of photons, $\Omega_{X}$ is the density of cold dark matter, and ${\rm h}=H_0/(100~{\rm km~s^{-1}~Mpc^{-1}})$.  We choose $\Omega_{X}=1$ for our analysis.  

\begin{figure}
\begin{center}
\epsfig{file=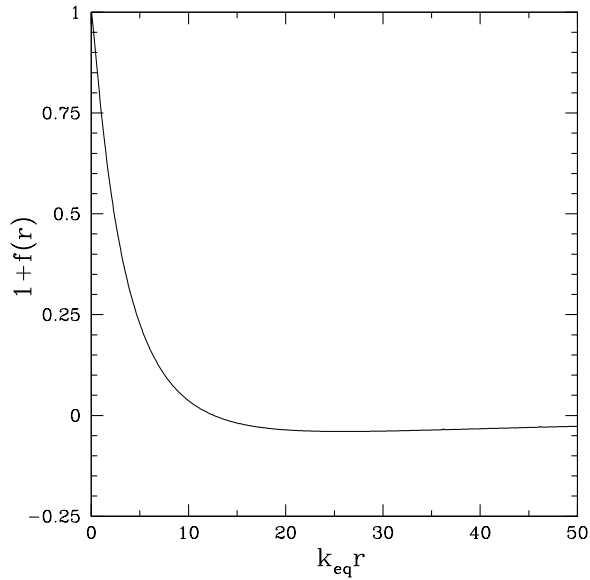,width=\picwidth,height =8cm}
\caption{The function $1+f(r)$ plotted versus $k_{eq}r$, where $k_{eq}$ is the wavenumber of the dominant perturbation mode.}  
\label{twopoint}
\end{center}
\end{figure}

Using this spectrum, 
\be
\langle \left(\nabla\Phi_{o(1)}\right)^2 \rangle=\frac{9C^2k_{eq}^2}{4}\int^{\infty}_0 ydyT^2(y)~,
\ee
where $k_{eq}=1/\lambda_c=\Omega_{X}{\rm h}^2\theta^{-1/2}{\rm Mpc}^{-1}\approx 3000\Omega_{X}{\rm h}\theta^{-1/2}H_0$ and the integral is approximately $2.31\times 10^{-2}$, using the transfer function in Eq. (\ref{transfer}).  So we finally find
\be
\langle \left(\nabla\Phi_{o(1)}\right)^2 \rangle\approx 9\times 10^{-6}H_0^2\left[\left(\frac{\Omega_{X}}{0.27}\right)\left(\frac{{\rm h}}{0.7}\right)\right]^{2}\theta^{-1}
\ee
and therefore
\be
\langle v^2_{o(1)} \rangle \approx 3\times 10^{-6}\left[\left(\frac{\Omega_{X}}{0.27}\right)\left(\frac{{\rm h}}{0.7}\right)\right]^{2}\theta^{-1}~.
\ee
Using the power spectrum (\ref{HZ}) we also find
\be
\left\langle {\bf n}\cdot{\bf v}({\bf r}_0,t){\bf n}\cdot{\bf v}({\bf r}_0+r{\bf n},t) \right\rangle=\frac{C^2k_{eq}^2}{H_0^2}\int_0^{\infty}ydyT^2(y)\left[\frac{1}{3}j_0\left(\frac{k_{eq}zy}{H_0}\right)-\frac{2}{3}j_2\left(\frac{k_{eq}zy}{H_0}\right)\right]~,
\label{nvnv}
\ee
where $j_0$ and $j_2$ are spherical Bessel functions of the first kind, defined in Eqs. (\ref{bess0}) and (\ref{bess2}) of Appendix C.  We plot $1+f(r)$, found by combining Eqs. (\ref{feqn}) and (\ref{nvnv}), in Figure~\ref{twopoint}.  Note that this becomes negative for $k_{eq}r\gtrsim 10$.  Note also that we have not used any truncation of the power on scales that are nonlinear.  If we instead were to impose a high-$k$ cutoff, so as not to include the effects of any modes that have $\Delta^2(k)>1$, then this would lead to differences of a factor of about two.  A different approach would be to include the quasi-linear regime, with the power spectrum given from N-body simulations \cite{Smith}.

\begin{figure}
\begin{center}
\epsfig{file=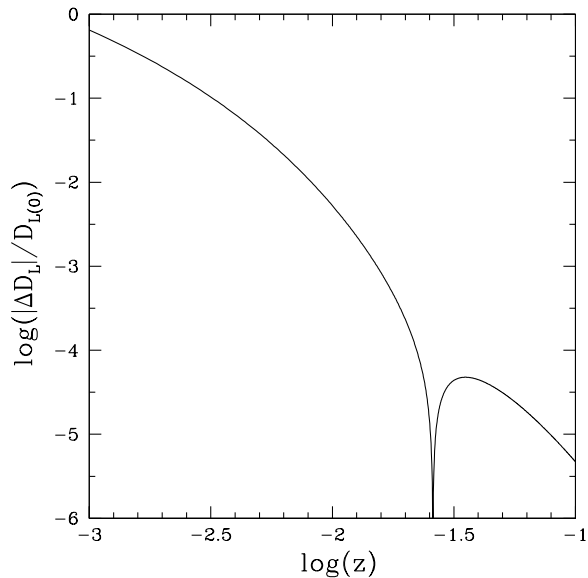,width=\picwidth,height =8cm}
\caption{The relative size of the perturbation $\log[|\Delta D_L(z)|/D_{L(0)}(z)]$ plotted versus $\log(z)$, assuming that the dominant perturbation wavelength is $10^{3}$ times smaller than the Hubble scale: $k_{eq}/H_0=10^{3}$.}  
\label{perts}
\end{center}
\end{figure}

We will specialize to $k_{eq}/H_0=1000$ for the rest of this paper, which yields
\be
\langle v^2_{o(1)} \rangle \approx 8.34\times 10^{-6}~.
\ee
In Figure~\ref{perts}, we show how the perturbation $\Delta D_L(z)$ scales relative to the unperturbed luminosity distance $D_{L(0)}(z)$, for the choice $k_{eq}/H_0=10^{3}$.
Note that we are plotting the logarithm of the absolute value, as the perturbation changes sign from positive to negative as one looks at larger distances.  By inspection, it becomes clear that $\Delta D_L(z)$ is not actually a perturbation for very small redshifts, i.e. for where $|\Delta D_{L}|/D_{L(0)}\sim 1$, and thus our computation of $\Delta D_L$ is no longer valid in that regime.  Indeed, it is well known that the peculiar velocities of objects within the Local Supercluster are not small when compared to their redshifts.  However, this will not be a problem in practice, as Type Ia supernovae at such small redshifts are typically not used for cosmological parameter fitting.  We will eventually take this breakdown of perturbation theory into account by imposing a lower cutoff $z_{min}$ when we fit our data to a theoretical model.  By eye, we see that it should be safe to choose $z_{min}\sim 0.01$.

\section{The perturbation to the inferred cosmological constant}

\subsection{Finding the best-fit FRW model}

We may now find the inferred cosmological constant and deceleration parameter by analyzing Eq. (\ref{finalanswer}) within the context of what one would expect in a homogeneous model.  The lowest order perturbation to the luminosity distance depends on the difference between the peculiar velocities at the source and at the observer, and so the question that we now ask is: How do peculiar velocities and their correlations affect inferences drawn from data about cosmological models?  We cannot simply Taylor expand Eq. (\ref{finalanswer}) around the observer to find $q_0$.  This is because $f$ varies on short lengthscales of order $k_{eq}^{-1}\sim 10~{\rm Mpc}$, so that a Taylor series expansion would effectively mean computing $q_0$ from $D_L(z)$ within this unrealistically short lengthscale.  A good alternative then is to fit the perturbed luminosity distance over a finite range of redshifts to what one would expect in a homogeneous model with matter and a cosmological constant.

Suppose that the observer can measure
redshifts $\{z_i\}$ for a set of distant objects arbitrarily
well.  From the distance determinations $\{D_{Li}\}$, the observer can compute
$\{r_i=D_{Li}/(1+z_i)\}$, and we can therefore take $\{z_i,r_i\}$ to
be the data gathered by the observer. Suppose also that in actuality the Universe is spatially flat with
Hubble parameter $H_0$ and matter only. Let
\be
r_i=H_0^{-1}\left[F(z_i)+\di(z_i)\right]
\ee
be the physical value of $r_i$, where for a flat matter-only cosmology
\be
F(z_i)=\int_0^{z_i}{dz\over(1+z)^{3/2}}=2\left[1-{1\over\sqrt{1+z_i}}
\right]
\label{Feqn}
\ee
and $\di(z_i)$ (not to be confused with the matter perturbation power spectrum) is the non-FRW contribution to $r_i$, from fluctuations via velocity differences.  From Eq. (\ref{deltalow}), we find the ensemble averaged perturbation
\ba
\di(z_i)&\approx&\frac{C^2k_{eq}^3}{H_0^3}\int_0^{\infty}dyT^2(y)\Bigg[\frac{y\cos(k_{eq}z_iy/H_0)}{k_{eq}z_i/H_0}-3\frac{\sin(k_{eq}z_iy/H_0)}{(k_{eq}z_i/H_0)^2}-6\frac{\cos(k_{eq}z_iy/H_0)}{(k_{eq}z_i/H_0)^3y}\nonumber\\& &
\qquad\qquad\qquad\qquad~~ +6\frac{\sin(k_{eq}z_iy/H_0)}{(k_{eq}z_i/H_0)^4y^2}\Bigg]~.
\label{deltafull}
\ea

The observer fits the data to a FRW model that is slightly curved
and has a small cosmological constant. The fitted model is then
\be
r_i^{fit}=\int_0^{z_i}{dz\over H(z)}-{k\over 6}\left(\int_0^{z_i}{dz\over
H(z)}\right)^3~,
\ee
where $k=(\omm+\oml-1)\hfit^2$ and 
\ba
H^2(z)&=&\hfit^2\left[\omm(1+z)^3+(1-\omm-\oml)(1+z)^2+\oml\right]
\nonumber\\
&=&\hfit^2(1+z)^3\left[1-{(1-\omm)z\over 1+z}-{\oml z(2+z)\over
(1+z)^3}\right]~;
\ea
here $\hfit$ is the fitted Hubble parameter, and $\omm$ and
$\oml$ are the density parameters for matter and for the cosmological
constant, respectively. Let us work to first order in $1-\omm$ and $\oml$,
a simplification which ought to suffice as
long as $\di\ll 1$. Thus, the fitted model is
\ba
r_i^{fit}&=&\hfit^{-1}\left[F(z_i)+(1-\omm)G(z_i)+\oml I(z_i)\right]
\nonumber\\
&\equiv&\hfit^{-1}\left[F(z_i)+\epsm G(z_i)+\epsl I(z_i)\right]~,
\ea
where $F(z_i)$ is the same as before, and we have defined
\be
G(z)={1\over 2}\int_0^z{dz\,z\over(1+z)^{5/2}}+{1\over 6}[F(z)]^3
\label{Geqn}
\ee
and
\be
I(z)={1\over 2}\int_0^z{dz\,z(2+z)\over (1+z)^{9/2}}
-{1\over 6}[F(z)]^3~.
\label{Ieqn}
\ee
There are three fitting parameters: $\hfit$, $\epsm=1-\omm$
and $\epsl=\oml$.

From the data and our model we can compute a likelihood 
function. Assuming Gaussian uncertainties this will be
the exponential of
\ba
\chit&=&-{1\over 2}\sum_i{\left[r_i-r_i^{fit}(z_i)\right]^2\over\sigi}
\nonumber\\
&=&-{1\over 2}\sum_i{\left[\left(H_0^{-1}-\hfit^{-1}\right)F_i
+H_0^{-1}\di-\hfit^{-1}(\epsm G_i+\epsl I_i)\right]^2\over\sigi}~,
\ea
where $\sigma_i$ is the estimated uncertainty in the value
of $r_i$ inferred from observations and $Q_i\equiv Q(z_i)$ for $Q=F,G,I$. 

The next step is to maximize $\chit$ with respect to the 
parameters of the fit, which will
lead to a set of coupled nonlinear equations. To simplify, let
us linearize in the small parameters $\epsm$, $\epsl$, $\{\di\}$
and $h=\hfit/H_0-1$.  The resulting equations are
\be
\dfav=\epsm\gfav+\epsl\ifav-h\ffav~,
\ee
\be
\dgav=\epsm\ggav+\epsl\igav-h\gfav~,
\ee
and
\be
\diav=\epsm\igav+\epsl\iiav-h\ifav~,
\ee
where we have defined the average $\langle Q_i\rangle\equiv\sum_i Q_i/(N\sigi)$. Solving for the parameters of the fit, we get
\ba
\epsl&=&{\cal D}^{-1}\biggl[\dfav(\ifav\ggav-\igav\gfav)+
\dgav(\igav\ffav-\ifav\gfav)\nonumber\\& &
+\diav(\gfav^2-\ggav\ffav)\biggr]~,
\label{epsleqn}
\ea
\ba
\epsm&=&{\cal D}^{-1}\biggl[\dfav(\gfav\iiav-\ifav\igav)+
\dgav(\ifav^2-\iiav\ffav)\nonumber\\& &
+\diav(\igav\ffav-\gfav\ifav)\biggr]~,
\label{epsmeqn}
\ea
and
\ba
h&=&-{\cal D}^{-1}\biggl[\dfav(\igav^2-\iiav\ggav)+
\dgav(\gfav\iiav-\igav\ifav)\nonumber\\& &
+\diav(\ifav\ggav-\gfav\igav)\biggr]~,
\label{heqn}
\ea
where
\be
{\cal D}=\igav^2\ffav-2\ifav\igav\gfav-\iiav\ffav\ggav+\iiav\gfav^2
+\ifav^2\ggav~.
\label{Deqn}
\ee
These are fairly general for small $\di$, and show that there may
be contributions to $\epsl$, $\epsm$, and $h$ from velocity
fluctuations. 

Next, we need to compute the averages. To do this, we
recall that $F$ corresponds to comoving radial coordinate, modulo
a factor of $H_0^{-1}$. To the order of approximation underlying
our calculations, we can take the comoving source density to be
uniform. Moreover, we do not need to worry about Malmquist bias,
at least for Type Ia supernovae, which are very bright. Let us also assume
that all of the $\{\sigi\}$ are the same, to keep the problem
as simple as possible. Then $\sigi$ drops out of our expressions
for $\epsm$, $\epsl$, and $h$, although it remains in their uncertainties. We suppose that our source
catalog extends to some maximum value $F_{max}$, with a corresponding maximum redshift $z_{max}$. It is worth
remembering that $F<2$ is an absolute upper bound, and that
for $z<1$, $F<2-\sqrt{2}\approx 0.6$, so we will be dealing
with relatively small values of $F$ typically.  Moreover, as we have already noted in Figure~\ref{perts}, our small $\Delta_i$ assumption breaks down below a minimum redshift $z_{min}\lesssim 0.01$, but this is not a problem as no supernovae below this redshift have ever been used for cosmological model fitting \cite{Riess, Perlmutter}.  So we will assume a lower cutoff for all of our sums of $F_{min}$.  Then, for example,
\be
\ffav=\frac{3}{F_{max}^3-F_{min}^3}\int_{F_{min}}^{F_{max}} dF\,F^2\,F^2~,
\ee
and Eqs. (\ref{Feqn}), (\ref{Geqn}), (\ref{Ieqn}), and (\ref{Deqn}) give the lowest order result, assuming that $F_{max}^3\gg F_{min}^3$,
\be
{\cal D}\approx-\frac{1}{5268480}F_{max}^{12}~.
\ee

Keeping only lowest order terms in $F_{max}$ in the numerators of Eqs. (\ref{epsleqn}), (\ref{epsmeqn}), and (\ref{heqn}) as well, we get
\be
\epsl\approx -\frac{5268480}{16}\left[\frac{3\dfav}{784F_{max}^4}-\frac{3\dffav}{280F_{max}^5}+\frac{\dfffav}{140F_{max}^6}\right]~,
\label{epsleqn2}
\ee
\be
\epsm\approx-2\epsl~,
\label{correspondence}
\ee
and
\be
h\approx -\frac{5268480}{16}\left[\frac{\dfav}{448F_{max}^2}-\frac{\dffav}{168F_{max}^3}+\frac{3\dfffav}{784F_{max}^4}\right]~.
\ee
We see that if $\di\propto F_i$, then $\epsl$ is zero, because the three terms in Eq. (\ref{epsleqn2}) cancel. This means that if $\di$ arises from velocity
correlations, it is only the correlation function of velocities
at two separated points that matters, not the
RMS velocity at a point.  Also note that, for this fitting procedure, the deceleration parameter is still $q_0=1/2$, since
\ba
\Delta q_0&=&q_0-{1\over 2}=
-{1\over 2}-(\ddot a a/H^2)_0\nonumber\\
&=&{1\over 2}(\omm-1-2\oml)=
{1\over 2}(-\epsm-2\epsl)={1\over 2}(2\epsl-2\epsl)=0~
\ea
from Eq. (\ref{correspondence}), in agreement with Refs. \cite{Flanagan}, \cite{Hirata}, and \cite{VFW}.

The perturbation $\di$, given in Eq. (\ref{deltafull}), depends on the correlation function $f(r)$, and so it does contribute to $\epsl$. For $z_{min}=0.02$ and $z_{max}=0.15$, we numerically integrate to find that the best-fit cosmological constant density is $\Omega_{\Lambda}\approx 0.004$.  Table~\ref{lambda} gives a few more results for the best-fit values for $\epsl$, $\epsm$, and $h$ as a function of the two limiting redshifts $z_{min}$ and $z_{max}$ in the continuum limit, where we have made the assumption that the number of sources $N$ is very large: $N\rightarrow\infty$. In this limit, $\di(z_i)\rightarrow\Delta(z)$ and
\be
\epsl=\int_{F_{min}}^{F_{max}}dF w(F)\Delta(F)~,
\ee
where we have the weighting function
\be
w\left(F\right)\equiv-\frac{5268480}{16}\left(\frac{3F}{784F_{max}^4}-\frac{3F^2}{280F_{max}^5}+\frac{F^3}{140F_{max}^6}\right)~.
\ee
We also plot these results in Figure~\ref{omega}, in the Introduction. Note that $\Omega_{\Lambda}$ may be positive or negative, depending on the redshift range, since $\Delta D_L$ changes sign in the region of interest.
\begin{table}[h]
\caption{Best-fit parameters in the continuum limit for a few values of the source catalog limiting redshifts $z_{min}$ and $z_{max}$, also for the choice that the dominant perturbation wavelength is $10^{3}$ times smaller than the Hubble scale: $k_{eq}/H_0=10^{3}$.}
\begin{ruledtabular}
\begin{tabular}{ccccc}
$z_{min}$ & $z_{max}$ & $\Omega_{\Lambda}$ & $1-\Omega_{M}$ & $H_{fit}/H_0-1$ \\
\hline
0.01 & 0.1 & $-0.018$ & 0.036 & $-4.3\times 10^{-5}$ \\
~ & 0.2 & $0.0016$ & $-0.0032$ & $4.0\times 10^{-5}$ \\
\hline
0.03 & 0.1 & $0.0037$ & $-0.0074$ & $7.1\times 10^{-5}$ \\
~ & 0.2 & $0.0020$ & $-0.0040$ & $4.7\times 10^{-5}$
\end{tabular}
\end{ruledtabular}
\label{lambda}
\end{table}

In order to test the robustness of these continuum limit calculations, we have also applied our fitting procedure to randomly-generated catalogs of synthetic redshift data.  To generate a data point $F_i$ for such a catalog, we assume that the quantity $(F_i^3-F_{min}^3)/(F_{max}^3-F_{min}^3)$ is distributed uniformly between 0 and 1.  In this way, we create catalogs of $N=100$ data points, wherein each data point is a value of $F_i$ for a source with a random location.  For each data point, we use the ensemble averaged formula for $\Delta D_L(z)$ to find $\Delta_i$.  We then fit these data to a homogeneous model as outlined above, using sums instead of integrals.  Using 20 randomly-generated catalogs, the average best-fit values for $\Omega_{\Lambda}$ are summarized in Table~\ref{lambda2}, along with their standard deviations.  We also found the best-fit cosmological constant with 50 catalogs for $z_{min}=0.02$ and $z_{max}=0.15$, to find $\Omega_{\Lambda}=0.005\pm 0.001 $.
\begin{table}[h]
\caption{Best-fit parameters for 20 catalogs of N=100 samples each, for a few values of the source catalog limiting redshifts $z_{min}$ and $z_{max}$.  We have also made the choice that the dominant perturbation wavelength is $10^{3}$ times smaller than the Hubble scale: $k_{eq}/H_0=10^{3}$.}
\begin{ruledtabular}
\begin{tabular}{ccc}
$z_{min}$ & $z_{max}$ & $\Omega_{\Lambda}$ \\
\hline
0.01 & 0.1 & $-0.020 \pm 0.002$ \\
~ & 0.2 & $0.002 \pm 0.001$ \\
\hline
0.03 & 0.1 & $0.014 \pm 0.001$ \\
~ & 0.2 & $0.0025 \pm 0.0004$
\end{tabular}
\end{ruledtabular}
\label{lambda2}
\end{table}

\subsection{Variance}

Although the best-fit values for $\Omega_{\Lambda}$ of the previous subsection are very small, we must keep in mind that they are derived from the ensemble averaged perturbation to the luminosity distance.  For a given source, this ensemble averaged perturbation will be far smaller than the leading order perturbation, which depends linearly on the peculiar velocity.  This linear perturbation will be the main source of the variance in the best-fit parameters, and this variance should overwhelm the systematic error for typical supernova sample sizes.  This complication was pointed out by Ref. \cite{HG} and it was shown to cause errors of $\Delta\Omega_{\Lambda}\approx -0.04$ for a sample of actual nearby supernovae in Ref. \cite{Velocities}.

Consider our expression for the best-fit $\Omega_{\Lambda}$, in terms of N discrete sources, rewritten as a weighted sum,
\be
\Omega_{\Lambda}=\frac{1}{N}\sum_i w\left(F_i\right)\Delta_i~.
\ee
What we have computed is the ensemble average of this,
\be
\langle\Omega_{\Lambda}\rangle=\frac{1}{N}\sum_i w\left(F_i\right)\langle\Delta_i\rangle~.
\ee
The variance is then
\ba
\sigma_{\Lambda}^2&=&\left\langle \left(\Omega_{\Lambda}-\langle\Omega_{\Lambda}\rangle\right)^2 \right\rangle=\langle\Omega_{\Lambda}^2\rangle+O\left(\delta^3\right)\nonumber\\
&=&\frac{1}{N^2}\sum_{i,j}w(F_i)w(F_j)\langle \Delta_i\Delta_j \rangle~,
\ea
which has two types of terms contributing: those with $i=j$ and those with $i\neq j$.  Separating these, we have $\sigma_{\Lambda}^2=\sigma_1^2+\sigma_2^2$, where
\be
\sigma_1^2\equiv\frac{1}{N^2}\sum_{i}w^2(F_i)\langle \Delta_i^2 \rangle
\ee
and
\be
\sigma_2^2\equiv\frac{1}{N^2}\sum_{i\neq j}w(F_i)w(F_j)\langle \Delta_i\Delta_j \rangle ~.
\label{VAR2}
\ee

In the continuum limit $N\rightarrow\infty$, the first piece of the variance becomes
\be
\sigma_1^2\approx \frac{1}{N}\frac{3}{F_{max}^3}\int_0^{F_{max}}F^2dFw^2(F)\langle \Delta^2(F) \rangle
\label{contvar}
\ee
where, from Eq. (\ref{deltafull}),
\be
\langle \Delta^2(F) \rangle=\langle \Delta^2(H_0 r) \rangle =\left\langle {\bf n}\cdot\left[{\bf v}({\bf r})-{\bf v}(0)\right]{\bf n}\cdot\left[{\bf v}({\bf r})-{\bf v}(0)\right] \right\rangle\sim \langle v_o^2 \rangle~.
\ee
The integrand in Eq. (\ref{contvar}) is integrable as $F\rightarrow 0$, and so the quantity $\sigma_1$ is to a good approximation independent of $z_{min}$ for small $z_{min}$.  Thus we can for simplicity take $z_{min}=0$. After integrating, we find
\be
\sigma_1^2\sim\frac{100}{N}\left(\frac{\langle v_o^2 \rangle}{8\times 10^{-6}}\right)\left(\frac{z_{max}}{0.2}\right)^{-6}~.
\label{var1}
\ee
For a source catalog of 100 sources out to a limiting redshift $z_{max}=0.2$, we find that this variance is significant: $\sigma_1^2\sim 1$.

The second piece (\ref{VAR2}) of the variance does not depend on the sample size, although it does depend on $F_{max}$.  In the continuum limit,
\be
\sigma_2^2\approx \frac{9}{F_{max}^6}\int_0^{F_{max}}F^2dFw(F)\int_0^{F_{max}}\left(F'\right)^2dF'w(F')\langle \Delta(F)\Delta(F') \rangle
\label{sigma2}
\ee
where
\be
\langle \Delta(F)\Delta(F') \rangle=\frac{1}{3}\langle v_{o(1)}^2\rangle\left[f\left(\frac{F}{H_0}-\frac{F'}{H_0}\right)-f\left(\frac{F}{H_0}\right)-f\left(\frac{F'}{H_0}\right)\right]~.
\label{deltaffp}
\ee
Plugging Eq. (\ref{deltaffp}) into Eq. (\ref{sigma2}), then using Eqs. (\ref{feqn}) and (\ref{nvnv}), and then finally doing some rearranging, we find
\be
\sigma_2^2\approx \left(\frac{246960 CH_0}{F_{max}^5k_{eq}}\right)^2\int_0^y\frac{dy}{y^3}T^2(y)\left[I\left(\frac{2k_{eq}F_{max}}{H_0}y\right)\right]^2
\ee
where
\be
I(q)\equiv\int_0^1dx\left(\frac{3}{784}x-\frac{3}{280}x^2+\frac{1}{140}x^3\right)\left(\sin qx-qx\cos qx\right)~.
\ee
This result for $\sigma_2^2$ does not depend on the sample size, as it only depends on the size of the redshift range $F_{max}$, making it a measure of cosmic variance.  By integrating numerically, we find that it scales roughly as $F_{max}^{-8}$ and
\be
\sigma_2^2\sim 0.03\left(\frac{z_{max}}{0.2}\right)^{-8}~.
\label{var2}
\ee
For comparison, Ref. \cite{Velocities} uses a sample of 115 supernovae up to a redshift $z_{max}=1.01$, and they find an error from the data of $\Delta\Omega_{\Lambda}=-0.04$.  For this same scenario, we estimate $|\Delta\Omega_{\Lambda}|\approx 0.01$, from the sum of Eqs. (\ref{var1}) and (\ref{var2}).

\section{Consistency with prior results}

The method of analysis that we have presented in the previous sections differs from that of Refs. \cite{Rasanen, Notari, Kolb1, Kolb2}.  This is because of (i) a difference in gauge choice and (ii) a fundamental difference in the definition of what constitutes ``acceleration".  We have chosen to use the standard post-Newtonian gauge, and to define acceleration as being based on fitting the luminosity distance-redshift relation to that of a homogeneous model containing dust and a cosmological constant.  As this definition of acceleration is based only on observable quantities, performing our calculation in other gauges gives us the same results.  

In contrast, Refs. \cite{Rasanen, Notari, Kolb1, Kolb2} calculate the cosmological expansion rate, averaged over a constant time slice.  The motivation for doing this comes from the spatially-averaged Friedmann equations, also called the Buchert equations \cite{Buchert}.  In particular, Ref. \cite{Kolb1} defines the effective coarse-grained scale factor $a_D$ in terms of the average matter density: $\langle\rho\rangle_D\propto a_D^{-3}$, where the angle brackets $\langle\rangle_D$, with subscript $D$, denote an average over a spatial hypersurface $D$ at a given time. Then Ref. \cite{Kolb2} defines the coarse-grained Hubble rate
\be
H_D=\frac{\dot{a}_D}{a_D}=\frac{1}{3}\langle\theta\rangle_D
\ee
and the effective deceleration parameter
\be
q=-\frac{\dot{H}_D}{H_D^2}-1~.
\label{qkolb}
\ee
These measures of acceleration are somewhat arbitrary since the deceleration parameter (\ref{qkolb})
depends on the spatial hypersurface over which one averages.  Refs. \cite{Rasanen, Notari, Kolb1, Kolb2} use constant time slices in the comoving synchronous gauge.  In this gauge, the perturbation to the
the expansion $\theta$ is related quite simply to the perturbations to the trace of the
connection; from Ref. \cite{Kolb1},
\be
\langle \theta_{(1)} \rangle_D = \frac{1}{a}\langle \Gamma^i_{ti(1)} \rangle_D~,
\label{kolbtrace}
\ee
and similarly for $\theta_{(2)}$.  Ref. \cite{Notari} claims that spatially averaged perturbations could become quite large, which implies that our perception of the expansion rate of the Universe is significantly affected by inhomogeneity.  The culprit is the appearance of terms in $\Gamma^i_{ti(2)}$ with large numbers of spatial gradients, which naturally appear in the synchronous gauge.  These higher derivative terms, which do not appear in our method above, lead to a perturbative instability, wherein terms higher order in perturbation theory do not get smaller as expected.

Although the results of the previous sections appear to differ from the claims of Refs. \cite{Rasanen, Notari, Kolb1, Kolb2}, in fact the large fitting effect claimed in those papers arises at a higher post-Newtonian order than we have computed.  In this section we show that our results are consistent with theirs to the order we have computed. Our method of computation could be extended to higher post-Newtonian order, which would allow for a detailed confrontation with their claims.

However, we believe that our result of a small fitting effect is robust, in the sense that it will not be altered by the inclusion of effects that are higher order in $\epsilon$ and/or $\delta$.  This belief is based on the structure of the post-Newtonian expansion of Einstein's equations, and on the fact
that we are computing a gauge-invariant observable.  If this is true, then our conclusion is in disagreement with Refs. \cite{Rasanen,Notari,Kolb1, Kolb2}.  

We believe the most likely reason for the
disagreement is that we compute a gauge-invariant observable that is
directly and uniquely related to supernova observations, whereas the
quantities computed in Refs. \cite{Rasanen, Notari,Kolb1, Kolb2} have
some arbitrariness and are not directly related to observations.
The proposal of Refs. \cite{Rasanen, Notari, Kolb1, Kolb2} that there
might be a large backreaction effect in terms of $q_D$ does not
necessarily imply that observers will measure large deviations from
FRW dynamics.  As mentioned above, spatially averaged perturbations are dependent on one's coordinate choice, in the sense that a constant time hypersurface in one coordinate system is most likely not going to be a constant time hypersurface in a different coordinate system.  These averages are unlikely to be directly observable, and are not uniquely related to the cosmic acceleration inferred from cosmological observations.  As Hirata and Seljak \cite{Hirata} remarked, we ``cannot cover the entire universe with astronomers so as to measure spatially averaged quantities" such as $H_D$.  It is possible that the measure of acceleration (\ref{qkolb}) could be large while the observed acceleration is small.

We now turn to showing consistency of our results with those of Refs. \cite{Rasanen, Notari, Kolb1, Kolb2} to the order we have computed. We take our metric (\ref{basicmetric}) and transform it from the post-Newtonian gauge to the synchronous gauge.  We then compute from the transformed metric the
perturbation to the Hubble rate.  The relative size of the difference between $H_D$ and the expected FRW value $H$ determines whether or not there will be a large fitting effect.  As an example, we will now compute the ratio
\be
\frac{H_D-H}{H}\equiv\frac{\Delta H}{H}=\frac{\langle \theta_{(1)}+\theta_{(2)} \rangle_D}{3H}
\label{kolb}
\ee
where the spatial average involves integrating with respect to the perturbed volume element $dV=\sqrt{g_{space}}d^3x$, where $g_{space}$ is the determinant of the spatial part of the metric.  Note that the quantity that we define as $\Delta H/H$ differs from what is computed in Refs. \cite{Rasanen, Notari, Kolb1, Kolb2}, although we do find the same qualitative result at the end of the day.  Below we show that this quantity is small to Newtonian order, in correspondence with what was found in \cite{Kolb1}, even though it involves a sum of terms that can be large individually. The reason these terms are large is that in synchronous coordinates metric perturbations can be of order $\delta$, which may be of considerable size even though there are no large gravitational potentials anywhere in the Universe. By contrast, in our calculation based on standard post-Newtonian coordinates, metric perturbations are at most of order  $\epsilon^2\delta$, which is always small. In this sense, perturbation expansions are much better behaved in the standard post-Newtonian coordinates than in synchronous coordinates.

We start by reviewing the transformation from standard post-Newtonian
coordinates (\ref{basicmetric}) to synchronous coordinates; a detailed
discussion is presented in Appendix D.
Begin with the second order perturbed FRW metric in the gauge
\be
ds^2=a^2(\eta)\left[-\left(1+2\Phi_{(1)}+2\Phi_{(2)}\right)d\eta^2+\left(1-2\Phi_{(1)}-2\Phi_{(2)}\right)\delta_{ij}dX^idX^j\right]~,
\ee
where we are now using conformal and Cartesian coordinates for simplicity, and we will only need to work to Newtonian order.  We can then define the new coordinates $\tau$ and $\tilde{x}^i$ by
\be
\eta=\tau\left[1-\frac{1}{3}\Phi_{(1)}-\frac{1}{5}\Phi_{(2)}+\frac{2\tau^2}{45}\left(\nabla\Phi_{(1)}\right)^2\right]+O\left(\tau_0\varepsilon^4\right)+O(\tau_0\delta^3)
\ee
and
\be
X^i=\tilde{x}^i-\frac{\tau^2}{6}\Phi_{(1),i}-\frac{\tau^2}{20}\Phi_{(2),i}+\frac{\tau^4}{120}\Phi_{(1),ij}\Phi_{(1),j}+O\left(\tilde{x}^i\varepsilon^2\right)+O(\tilde{x}^i\delta^3)~,
\ee
where these potentials are fixed physical quantities, evaluated at $(\tau,\tilde{x}^i)$, and these spatial derivatives are in terms of the new coordinates.  We are also assuming that we have the growing mode only, for which we have the power law scalings $\Phi_{(1)}\propto \tau^0$ and $\Phi_{(2)}\propto\tau^2$. Then the line element becomes, to lowest order in $\varepsilon$, 
\ba
ds^2&=&a^2(\tau)\left[-d\tau^2+\tilde{g}_{ij}d\tilde{x}^id\tilde{x}^j\right]\nonumber\\
&=&a^2(\tau)\left\{-d\tau^2+\left[\delta_{ij}-\frac{\tau^2}{3}\Phi_{(1),ij}-\frac{\tau^2}{10}\Phi_{(2),ij}+\frac{\tau^4}{60}\Phi_{(1),ijk}\Phi_{(1),k}+\frac{2\tau^4}{45}\Phi_{(1),ik}\Phi_{(1),jk}+O\left(\varepsilon^2\right)\right]d\tilde{x}^id\tilde{x}^j\right\}~,
\ea
which is now in a synchronous gauge. Note that the metric now has perturbations of order $\varepsilon^0\delta\sim \delta$.  These order $\delta$ perturbations will lead to the appearance of large terms in $\Delta H/H$, which will cancel when averaged.  Then we find
\be
\sqrt{g_{space}}=a^3(\tau)\left[1-\frac{\tau^2}{6}\nabla^2\Phi_{(1)}+O\left(\delta^2\right)\right]~.
\ee
The spatial trace of the connection is
\be
\Gamma^i_{\tau i}=\frac{1}{2a^2}\tilde{g}^{ij}\left(a^2 \tilde{g}_{ij}\right)_{,\tau}
\ee
which receives the first and second order perturbations
\be
\Gamma^i_{\tau i(1)}=a(\tau)\theta_{(1)}=\frac{1}{2}\delta^{ij}\tilde{g}_{ij(1),\tau}=-\frac{\tau}{3}\nabla^2\Phi_{(1)}+O\left(\delta\varepsilon^2\right)
\ee
and
\ba
\Gamma^i_{\tau i(2)}=a(\tau)\theta_{(2)}&=&\frac{1}{2}\tilde{g}^{ij(1)}\tilde{g}_{ij(1),\tau}+\frac{1}{2}\delta^{ij}\tilde{g}_{ij(2),\tau}\nonumber\\
&=&-\frac{\tau^3}{45}\Phi_{(1),ij}\Phi_{(1),ij}-\frac{\tau}{10}\nabla^2\Phi_{(2)}+\frac{\tau^3}{30}\left(\nabla^2\Phi_{(1)}\right)_{,k}\Phi_{(1),k}+O(\delta^2\varepsilon^2)~.
\ea
Using the Fourier transformation (\ref{ft}), taking an ensemble average, and using the result that $\langle\nabla^2\Phi_{(2)}\rangle=0$ (see Appendix C), we find from Eqs. (\ref{kolb}) and (\ref{kolbtrace})
\ba
\frac{\Delta H}{H}&\approx&\frac{1}{3Ha}\left\langle\frac{\tau^3}{18}\left(\nabla^2\Phi_{(1)}\right)^2-\frac{\tau^3}{45}\Phi_{(1),ij}\Phi_{(1),ij}+\frac{\tau^3}{30}\left(\nabla^2\Phi_{(1)}\right)_{,k}\Phi_{(1),k}\right\rangle\nonumber\\
&=&\frac{\tau^3}{135Ha}\left\langle\left(\nabla^2\Phi_{(1)}\right)^2-\Phi_{(1),ij}\Phi_{(1),ij}\right\rangle\nonumber\\
&=&\frac{\tau^3}{135Ha}\left\langle\left[\Phi_{(1),i}\nabla^2\Phi_{(1)}-\Phi_{(1),j}\Phi_{(1),ij}\right]_{,i}\right\rangle~,
\label{kolb2}
\ea
which is consistent with the lowest order result of Ref. \cite{Kolb1}. This spatial average is a boundary term, whose ensemble average vanishes.

Although (\ref{kolb2}) vanishes, it contains terms with two more powers of $k/H_0$ than what one would find in the post-Newtonian gauge.  It is these terms that Refs. \cite{Kolb2,Notari} argue will lead to a large effect at higher order in perturbation theory.  In other words, using the synchronous gauge and defining acceleration in terms of spatially averaged expansion parameters can lead to a conceivably large correction.  This is in contrast to our earlier method, wherein we calculate the observable effect, which is very small.  Note that our expansion (\ref{dllong}) for $D_L(z)$ contains no four-derivative terms like those in (\ref{kolb2}).

\section{Conclusions}

We have computed the inhomogeneity-induced perturbations to the redshifts and luminosity distances that a comoving observer would measure to first post-Newtonian order, i.e. we have computed $z$ and $H_0D_L$ to order $\varepsilon^3\sim(v/c)^3$, and to second order in the density perturbation $\delta=(\rho-\langle\rho\rangle)/\langle\rho\rangle$.  Assuming a flat and matter-dominated background cosmology, the perturbed luminosity distance-redshift relation is given by Eq. (\ref{finalanswer}). The perturbations to $D_L(z)$ depend on the correlation between the peculiar velocities at the observer and at the source. Roughly speaking, these perturbations are of order $\Delta D_L/D_L\sim 10^{-5}$ when $z\sim 0.1$.  The luminosity distance-redshift relation was averaged over viewing angles and over an ensemble of realizations of the density perturbation.  The result is gauge invariant, as it corresponds to a measurable quantity. We then fit this function to what one would expect in a homogeneous FRW cosmology, containing dust and a cosmological constant, to deduce the corresponding perturbation to the inferred cosmological constant density.

The inferred $\Omega_{\Lambda}$ depends on the limiting redshifts $z_{min}$ and $z_{max}$ of the sample, and we summarize the best-fit values of $\Omega_{\Lambda}$ for different values of these limiting redshifts in Figure~\ref{omega} and Table~\ref{lambda}.  These ensemble averaged results indicate that we are justified in fitting low-$z$ supernova data to homogeneous models, as long as we use supernova data that spans a large enough redshift range.  For instance, assuming that we have luminosities and redshifts from $z_{min}=0.02$ out to $z_{max}= 0.15$, the errors induced by the ``fitting problem" are small: $\Omega_{\Lambda}\sim 0.004$.  Such errors are not large enough to explain the measured value $\Omega_{\Lambda}\approx 0.7$.  This is what we would expect, since we have other evidence to suggest that our universe contains dark energy from large scale structure surveys, from the CMB power spectrum, and from weak lensing.  

In contrast to the small value of the best-fit $\Omega_{\Lambda}$ for the ensemble averaged luminosity distance-redshift relation, we find that relatively large errors are possible due to fluctuations in $D_L(z)$, specifically from terms that are linear in peculiar velocities.  This effect was noted in Ref. \cite{HG} and then calculated in Ref. \cite{Velocities} for an actual nearby supernova data set.  We find that the associated variance in $\Omega_{\Lambda}$ has two components, one that depends on the number of sources $N$, $\sigma_1^2\sim (100/N)(z_{max}/0.2)^{-6}$, and one that does not, $\sigma_2^2\sim 0.03(z_{max}/0.2)^{-8}$.

It should be stressed that our goal in this paper was only to find a rough estimate of the fitting effect.  One potential weakness of our analysis is that we have assumed that $\delta<1$, and thus we do not address the effects of highly nonlinear structures.  Such nonlinear modes could be included by using the full nonlinear power spectrum from N-body simulations \cite{Smith}, and we estimate that this would change the result by approximately a factor of two.  Furthermore, we have assumed that the observer is in a random location in the Universe, and has no knowledge of his/her own peculiar velocity.  One can redo the calculation for an observer who knows and corrects for this velocity.

It has been claimed that there exists a perturbative instability, where successive orders in an expansion in powers of $\delta$ do not get smaller \cite{Notari, Kolb1, Kolb2}.  We do not see any indications of such an instability with our method. When one defines ``acceleration" in terms of only directly observable quantities, as we did in Sections II through V, the fitting effect one obtains is small.

\begin{acknowledgments}
R.A.V. is supported by an American dissertation fellowship from the AAUW Educational Foundation.  This research was supported in part by NSF grants PHY-0457200 and PHY-0555216.  We also thank Syksy R\"{a}s\"{a}nen for pointing out an error in an earlier version of this manuscript.
\end{acknowledgments}

\appendix

\section{Combining the redshift and luminosity distance relations}

Adding the redshift equations (\ref{zzero}), (\ref{zfirst}), and (\ref{zsecond}) yields
\ba
z(\lambda,\theta,\phi)&=&\left[\frac{2r}{3t}+\frac{r^2}{9t^2}+\frac{4r^3}{27t^3}+O\left(\varepsilon^4\right)\right]\nonumber\\& &
+\Bigg[v^r_{s(1)}-v^r_{o(1)}+\Phi_{o(1)}-\Phi_{s(1)}+\frac{2r}{3t}\left(v^r_{s(1)}-v^r_{o(1)}\right)-2\int^r_0\dot{\Phi}_{(1)}dr'+\frac{2r}{3t}\Phi_{o(1)}\nonumber\\& &
-\frac{2r}{t}\Phi_{s(1)}-\frac{r^2}{9t^2}v^r_{o(1)}+\frac{r^2}{3t^2}v^r_{s(1)}+O\left(\varepsilon^4\delta\right)\Bigg]\nonumber\\& &
+\Bigg\{v^r_{s(2)}-v^r_{o(2)}+\Phi_{o(2)}-\Phi_{s(2)}+\frac{2r}{3t}\left(v^r_{s(2)}-v^r_{o(2)}\right)+\frac{1}{2}\left(v_{s(1)}^2-v_{o(1)}^2\right)+\left(v^r_{o(1)}\right)^2\nonumber\\& &
-v^r_{o(1)}v^r_{s(1)}-2\int^r_0\dot{\Phi}_{(2)}dr'+\left(v_{\theta(1)}k_{(1)}^{\theta}+v_{\phi(1)}k_{(1)}^{\phi}\right)_o-\left(v_{\theta(1)}k_{(1)}^{\theta}+v_{\phi(1)}k_{(1)}^{\phi}\right)_s+\frac{2r}{3t}\Phi_{o(2)}\nonumber\\& &
-\frac{2r}{t}\Phi_{s(2)}-\frac{r^2}{9t^2}v^r_{o(2)}+\frac{r^2}{3t^2}v^r_{s(2)}+\frac{r}{3t}\left[\left(v_{s(1)}\right)^2-\left(v_{o(1)}\right)^2\right]+\frac{2r}{3t}\left[\left(v^r_{s(1)}\right)^2+\left(v^r_{o(1)}\right)^2-v^r_{s(1)}v^r_{o(1)}\right]\nonumber\\& &
+\Phi_{o(1)}v^r_{o(1)}+\Phi_{s(1)}v^r_{o(1)}+\Phi_{o(1)}v^r_{s(1)}-3\Phi_{s(1)}v^r_{s(1)}+x_{(1)}^iv^r_{s(1),i}+O\left(\varepsilon^4\delta^2\right)\Bigg\}+O\left(\varepsilon\delta^3\right)~,
\label{zadded}
\ea
where the right hand side is evaluated at $r=r(\lambda)=-\lambda$ and $t=t(\lambda)=t_0+\lambda$. To point out a few of the above effects, the terms linear in velocity and linear in $\Phi$ correspond to the Doppler effect and the gravitational redshift, respectively.  We also see the second order Doppler shift with the $v^2$ terms, and the integrated Sachs-Wolfe effect with the integrated terms.  The perturbed luminosity distance is found from Eqs. (\ref{dlzero2}), (\ref{dlfirst}), and (\ref{dlsecond}) to be
\ba
D_L(\lambda,\theta,\phi)&=&\frac{\left(1+z\right)^2}{H_0}\frac{2r}{3t}\Bigg\{\left[1-\frac{r}{t}+\frac{8r^2}{9t^2}+O\left(\varepsilon^3\right)\right]\nonumber\\& &
-\left[\int^r_0\frac{dr'}{r'^2}\int^{r'}_0\left(r''\right)^2\nabla^2\Phi_{(1)} dr''+O\left(\varepsilon^3\delta\right)\right]\nonumber\\& &
-\left[\int^r_0\frac{dr'}{r'^2}\int^{r'}_0\left(r''\right)^2\nabla^2\Phi_{(2)} dr''+O\left(\varepsilon^3\delta^2\right)\right]+O\left(\varepsilon\delta^3\right)\Bigg\}~.
\label{DLadded}
\ea
Here we can see the effects of weak gravitational lensing.  Note that as the cosmological portion of the redshift goes to zero, and hence $r\rightarrow 0$, the luminosity distance also goes to zero, as expected.

By combining Eqs. (\ref{zadded}) and (\ref{DLadded}), we can eliminate $\lambda$ and compute $D_L$ as a function of $z$, $\theta$, and $\phi$. This computation can be carried out explicitly by using the fact that the expressions are power series in $\varepsilon$ and $\delta$.  This procedure gives:
\ba
D_L(z,\theta,\phi)&\approx&\frac{\left(1+z\right)^2}{H_0}\Bigg\{z-\frac{7}{4}z^2+\frac{19}{8}z^3+\left(-1+\frac{5}{2}z-\frac{33}{8}z^2\right)\left(v^r_{s(1)}+v^r_{s(2)}\right)\nonumber\\& &
+\left(1-\frac{5}{2}z+\frac{29}{8}z^2\right)\left(v^r_{o(1)}+v^r_{o(2)}\right)+\left(1-\frac{1}{2}z\right)\left(\Phi_{s(1)}+\Phi_{s(2)}\right)+\left(-1+\frac{5}{2}z\right)\left(\Phi_{o(1)}+\Phi_{o(2)}\right)\nonumber\\& &
+\left(\frac{1}{2}-\frac{5}{4}z\right)\left(v_{o(1)}^2-v_{s(1)}^2\right)+\left(-\frac{7}{4}+\frac{29}{8}z\right)\left(v^r_{o(1)}\right)^2+\left(-\frac{3}{4}+\frac{9}{8}z\right)\left(v^r_{s(1)}\right)^2+\left(\frac{5}{2}-\frac{23}{4}z\right)v^r_{o(1)}v^r_{s(1)}\nonumber\\& &
+\frac{1}{2}v^r_{o(1)}\left(\Phi_{o(1)}-\Phi_{s(1)}\right)+\frac{5}{2}v^r_{s(1)}\left(\Phi_{s(1)}-\Phi_{o(1)}\right)-x_{(1)}^iv^r_{s(1),i}+\int^r_0\left(\dot{\Phi}_{(1)}+\dot{\Phi}_{(2)}\right)dr'\nonumber\\& &
+\left(v_{\theta(1)}k_{(1)}^{\theta}+v_{\phi(1)}k_{(1)}^{\phi}\right)_s-\left(v_{\theta(1)}k_{(1)}^{\theta}+v_{\phi(1)}k_{(1)}^{\phi}\right)_o-\left(z+v^r_{o(1)}-v^r_{s(1)}\right)\int^r_0\frac{dr'}{r'^2}\int^{r'}_0\left(r''\right)^2\nabla^2\Phi_{(1)} dr''\nonumber\\& &
-z\int^r_0\frac{dr'}{r'^2}\int^{r'}_0\left(r''\right)^2\nabla^2\Phi_{(2)} dr''+\left(v^r_{s(1)}-v^r_{o(1)}\right)\left[z\frac{d}{dz}\int^r_0\frac{dr'}{r'^2}\int^{r'}_0\left(r''\right)^2\nabla^2\Phi_{(1)} dr''-2\frac{d}{dz}\int^r_0\dot{\Phi}_{(1)}dr'\right]\nonumber\\& &
+\left[\Phi_{s(1)}-\Phi_{o(1)}+\left(1+\frac{1}{2}z\right)\left(v^r_{o(1)}-v^r_{s(1)}\right)\right]\frac{d}{dz}\Phi_{s(1)}\nonumber\\& &
-\left[2\int^r_0\dot{\Phi}_{(1)}dr'+\left(-1+\frac{3}{2}z\right)\Phi_{o(1)}+\left(1+\frac{1}{2}z\right)\Phi_{s(1)}+\left(1-\frac{3}{2}z+\frac{13}{8}z^2\right)v^r_{o(1)}+\left(-1+\frac{3}{2}z-\frac{17}{8}z^2\right)v^r_{s(1)}\right]\nonumber\\& &
\times\frac{d}{dz}v^r_{s(1)}\Bigg\} ~,
\label{dllong}
\ea
where, to leading order, $d/dz\approx (3t_0/2)\partial/\partial r$.  The functions of $r$ and $t$ that appear on the right hand side of Eq. (\ref{dllong}) are evaluated at $r=z/H_0$ and $t=t_0-z/H_0$.  Note that the redshift $z$ here is the full redshift as measured by the observer.  Next we need to average $D_L(z,\theta,\phi)$ over viewing angles in the observer's rest frame, and also take an ensemble average.  In doing so, the averages of first order quantities will vanish.  We also will find that we will only need the second order velocities and potentials to Newtonian order, so that we may compute the lowest-order effect.

\section{Newtonian second-order perturbation theory}

In terms of comoving coordinates ${\bf r}={\bf x}/a(t)$ \cite{Peebles2}, the equations of Newtonian hydrodynamics are
\be
\frac{\partial\delta}{\partial t}+\frac{1}{a}{\bf \nabla}\cdot\left[\left(1+\delta\right){\bf v}_p\right]=0~,
\label{euler1}
\ee
\be
\frac{\partial{\bf v}_p}{\partial t}+\frac{\dot{a}}{a}{\bf v}_p+\frac{1}{a}\left({\bf v}_p\cdot{\bf \nabla}\right){\bf v}_p=-\frac{{\bf \nabla}\Phi_p}{a}~,
\label{euler2}
\ee
and
\be
\nabla^2\Phi_p=4\pi\rho_0 a^2\delta~,
\label{euler3}
\ee
where ${\bf v}_p={\bf v}_{(1)}+{\bf v}_{(2)}+\ldots$ is the peculiar velocity, $\Phi_p=\Phi_{(1)}+\Phi_{(2)}+\ldots$ is the perturbation to the Newtonian gravitational potential, the density contrast is $\delta=[\rho({\bf r},t)-\rho_0(t)]/\rho_0(t)$, and the zeroth order quantities are given in Section II.  The Newtonian first order results are very well known; for a detailed review, see Peebles \cite{Peebles2}.  For a Newtonian analysis to second order in $\delta$, see Ref. \cite{ZH}.  

The first order result is that the density contrast consists of mode that grows with time, and one that decays with time:
\be
\delta_{(1)}({\bf r},t)=f({\bf r})t^{2/3}+g({\bf r})t^{-1}~,
\label{modes}
\ee
where $f$ and $g$ are functions of the spatial coordinates.  We will only consider the growing mode. It is useful to rewrite the hydrodynamic equations in terms of their Fourier modes. Writing
\be
\delta = \int\frac{d^3k}{(2\pi)^3}\delta_{{\bf k}}e^{i{\bf k}\cdot{\bf r}}
\ee
and
\be
\Phi_p = \int\frac{d^3k}{(2\pi)^3}\Phi_{{\bf k}}e^{i{\bf k}\cdot{\bf r}}~,
\label{FT}
\ee
Eq. (\ref{euler3}) becomes
\be
k^2\Phi_{{\bf k}}=4\pi\rho_0 a^2\delta_{{\bf k}}~.
\label{fpoisson}
\ee
The second order density contrast is
\be
\delta_{(2)}=\frac{9t^4}{14a^4t_0^4}\left(\Phi_{(1),ij}\Phi_{(1),j}+\frac{5}{2}\nabla^2\Phi_{(1)}\Phi_{(1),i}\right)_{,i}~;
\label{sod}
\ee
this result came from perturbing Eqs. (\ref{euler1})-(\ref{euler3}) to second order and then solving these by using the first order solutions, Eqs. (\ref{modes}) and (\ref{FT}).  It can be seen that the expected value of $\delta_{(2)}$ vanishes by substituting the mode expansion of $\Phi_{(1)}$ into Eq. (\ref{sod}): $\langle \delta_{(2)} \rangle=0$.  We also see from Eq. (\ref{fpoisson}) that $\langle\Phi_p\rangle$ depends only on boundary conditions; we can choose to add overall constants to $\Phi$ at each order in $\delta$, and it is natural to choose these constants to satisfy $\langle\Phi_{(1)}\rangle=\langle\Phi_{(2)}\rangle=0$.

Assuming that we only have the growing mode solution of Eq. (\ref{modes}), we find that the first order peculiar velocity is related to the Newtonian potential,
\be
{\bf v}_{(1)}({\bf r},t)=-\frac{t}{a(t)}{\bf \nabla}\Phi_{(1)}=-t^{1/3}t_0^{2/3}{\bf \nabla}\Phi_{(1)}~.
\ee
This averages to zero but its square does not.  The second order velocity perturbation is
\be
v^i_{(2)}=-\frac{3t^3}{14a^3}\Phi_{(1),ij}\Phi_{(1),j}
\ee
which also averages to zero: $\langle v_{(2)} \rangle=0$.  Note that these averages are ensemble averages, not spatial averages.

\section{Averaging the luminosity distance-redshift relation}

Now we can scrutinize the terms of Eq. (\ref{dllong}), so that we may find their angular and ensemble averages.  Note that the angular averages will be performed with respect to the observer's angles $(\tilde{\theta},\tilde{\phi})$, and so we will need to use the Jacobian given in Eq. (\ref{Jacobian}).  The first three terms of Eq. (\ref{dllong}) only depend on the background cosmology, and are unchanged after averaging, and all terms that are to first order in $\delta$ will have a vanishing ensemble average.  As shown in Appendix B, terms that depend on $v^i_{(2)}$ and $\Phi_{(2)}$ also average to zero.

In addition, there are many terms that have vanishing ensemble averages because they contain an odd number of spatial derivatives of the potential, such as
\be
\left\langle v^r_{o(1)}\Phi_{o(1)} \right\rangle=\left\langle v^r_{s(1)}\Phi_{s(1)} \right\rangle=0~,
\ee
\be
\left\langle v^r_{s(1)}\frac{\partial}{\partial r}v^r_{s(1)} \right\rangle=0~,
\ee
\be
\left\langle x^i_{(1)}v^r_{s(1),i} \right\rangle=0~,
\ee
\be
\left\langle v^r_{s(1)}\frac{d}{dz}\int^r_0\dot{\Phi}_{(1)}dr' \right\rangle=0~,
\ee
et cetera.  We also find that
\be
\left\langle \left(v_{\theta(1)}k_{(1)}^{\theta}+v_{\phi(1)}k_{(1)}^{\phi}\right)_s-\left(v_{\theta(1)}k_{(1)}^{\theta}+v_{\phi(1)}k_{(1)}^{\phi}\right)_o \right\rangle\sim O\left(\varepsilon^4\right)~,
\ee
since $v_{\theta(1)}k_{(1)}^{\theta}\sim v_{\phi(1)}k_{(1)}^{\phi}\sim\varepsilon^3$, and taking the difference of the averages at the source and at the observer introduces another factor of $z\sim\varepsilon$.  

We can further rewrite the average $\langle v_{s(1)}^2 \rangle$ by exploiting the power law scaling $v_{(1)}^2\propto t^{2/3}$, to find
\ba
\langle v_{s(1)}^2 \rangle &=& \left\langle \left(t_0^{2/3}t^{1/3}\nabla\Phi_{(1)}\right)^2 \right\rangle \approx \left\langle \left(t_0^{2/3}\nabla\Phi_{(1)}\right)^2\right\rangle \left(t_0-r\right)^{2/3} \nonumber\\
&=& \langle v_{o(1)}^2 \rangle \left[1-z+{\cal O}(z^2)\right] ~.
\ea
We also use $\langle (v_{(1)}^r)^2 \rangle=\langle v_{(1)}^2 \rangle/3$, and introduce the two point correlation function $f(r)$,
\be
\langle v^r_{s(1)}v^r_{o(1)} \rangle=\frac{1}{3}\langle v_{o(1)}^2 \rangle\left(1-\frac{1}{2}z\right)\left[1+f(r)\right]~,
\ee
where $f(r)$ is defined by
\be
\left\langle {\bf n}\cdot{\bf v}({\bf r}_0,t){\bf n}\cdot{\bf v}({\bf r}_0+r{\bf n},t) \right\rangle=\frac{1}{3}\langle v_{o(1)}^2 \rangle\left[1+f(r)\right]~,
\ee
and ${\bf n}$ is a unit vector that defines the viewing direction.  

We can write this correlation function in terms of a more general correlation function $c_{ij}(r)$, using the Fourier transform of Eq. (\ref{FT}) and Eqs. (\ref{deltaeqn})-(\ref{yeqn}):
\be
\langle v_{o(1)}^2\rangle c_{ij}(r)\equiv\left\langle v_i({\bf r}_0,t_0)v_j({\bf r}_0+{\bf r},t_0) \right\rangle=\frac{H_0^2}{4\pi}\int_0^{\infty}\frac{d^3kk_ik_j\Delta^2(k)e^{-i{\bf k}\cdot{\bf r}}}{k^7}~.
\ee
This function can be rewritten as
\be
\langle v_{o(1)}^2\rangle c_{ij}(r)\equiv H_0^2\left[\frac{1}{3}A(r)\delta_{ij}+\frac{r_ir_j}{r^2}B(r)\right]~,
\ee
where
\be
A(r)=\frac{3}{8\pi}\int_0^{\infty}\frac{d^3k\Delta^2(k)}{k^5}\left[1-\left({\bf k}\cdot{\bf r}\right)^2\right]e^{-i{\bf k}\cdot{\bf r}}=\int_0^{\infty}\frac{dk\Delta^2(k)}{k^3}\left[j_0(kr)+j_2(kr)\right]
\ee
and
\be
B(r)=\frac{1}{8\pi}\int_0^{\infty}\frac{d^3k\Delta^2(k)}{k^5}\left[3\left({\bf k}\cdot{\bf r}\right)^2-1\right]e^{-i{\bf k}\cdot{\bf r}}=\int_0^{\infty}\frac{dk\Delta^2(k)}{k^3}\left[-j_2(kr)\right]~,
\ee
and where we are using spherical Bessel functions of the first kind:
\be
j_0(x)=\frac{\sin x}{x}
\label{bess0}
\ee
and
\be
j_2(x)=\left(\frac{3}{x^3}-\frac{1}{x}\right)\sin x-\frac{3}{x^2}\cos x~.
\label{bess2}
\ee
It follows that
\ba
\frac{1}{3}\langle v_{o(1)}^2\rangle\left[1+f(r)\right]&=& \langle v_{o(1)}^2\rangle n^in^jc_{ij}(r)=H_0^2\left[\frac{1}{3}A(r)+B(r)\right]\nonumber\\
&=&\frac{C^2k_{eq}^2}{H_0^2}\int_0^{\infty}ydyT^2(y)\left[\frac{1}{3}j_0\left(\frac{k_{eq}zy}{H_0}\right)-\frac{2}{3}j_2\left(\frac{k_{eq}zy}{H_0}\right)\right]~,
\ea
where $k_{eq}=1/\lambda_c\sim 10^3 H_0$.  We plot $1+f(r)$ in Figure~\ref{twopoint}; we see that it falls to approximately zero for $r\gg \lambda_c \sim 10~{\rm Mpc}$, and thus we do not expect it to be important when measuring the distances to supernovae at redshifts $z\sim 0.1$.  Note also that $f$ becomes negative for large enough $r$.

Using these simplifications, we finally get
\be
D_L(z)=\frac{z}{H_0}\left(1+\frac{1}{4}z-\frac{1}{8}z^2\right)+\Delta D^{rms}_L(z)+\Delta D^{corr}_L(z)~,
\ee
where $\Delta D^{rms}_L(z)$ is the perturbation that depends on RMS quantities at a given point, which vanishes:
\be
\Delta D^{rms}_L(z)=0~,
\ee
and $\Delta D^{corr}_L(z)$ is the perturbation that depends on $f$.  To subleading order, this is
\be
\Delta D^{corr}_L(z)\approx\frac{(1+z)^2\langle v^2_{o(1)}\rangle}{H_0} \left[\frac{3}{2}f\left(\frac{z}{H_0}\right)-\frac{1}{3H_0}f'\left(\frac{z}{H_0}\right)\left(1-2z\right)\right]+O\left(\frac{f\varepsilon^3\delta^2}{H_0}\right)~,
\ee
where the subleading terms are suppressed by a factor of $\lambda_cH_0$ or $z$.  We will only use the lowest order piece,
\ba
\Delta D_L(z)&=&\Delta D_L^{corr}(z)\approx-\frac{\langle v^2_{o(1)}\rangle}{3H_0^2}f'\left(\frac{z}{H_0}\right)\nonumber\\
&=&\frac{C^2k_{eq}^3}{H_0^4}\int_0^{\infty}dyT^2(y)\Bigg[\frac{y\cos(k_{eq}zy/H_0)}{k_{eq}z/H_0}-3\frac{\sin(k_{eq}zy/H_0)}{(k_{eq}z/H_0)^2}-6\frac{\cos(k_{eq}zy/H_0)}{(k_{eq}z/H_0)^3y}\nonumber\\& &
\qquad\qquad\qquad\qquad~~ +6\frac{\sin(k_{eq}zy/H_0)}{(k_{eq}z/H_0)^4y^2}\Bigg]~.
\ea

\section{Transforming from the standard post-Newtonian gauge to the synchronous gauge}

In the standard post-Newtonian gauge discussed in Section II, we can rewrite the metric in terms of conformal coordinates,
\be
ds^2=a^2(\eta)\left[-\left(1+2\Phi_{(1)}+2\Phi_{(2)}\right)d\eta^2+\left(1-2\Phi_{(1)}-2\Phi_{(2)}\right)\delta_{ij}dX^idX^j\right]~,
\ee
where we will only need this to Newtonian order, and now the scale factor is $a(\eta)=(\eta/\eta_0)^2$.  We will define $\eta_0\approx 3t_0$ to be the conformal time today. This new time coordinate is related to that of Sections II - V by
\be
\eta=3\left(\frac{t}{t_0}\right)^{-2/3}t\left[1-\frac{r^2}{9t^2}+O\left(\frac{r^4}{t^4}\right)\right]=\frac{3}{a}t+O\left(t\varepsilon^2\right)~,
\ee
and the radial coordinates are related by
\be
R=\left(\frac{t}{t_0}\right)^{-2/3}r\left[1+\frac{r^2}{9t^2}+O\left(\frac{r^4}{t^4}\right)\right]=\frac{r}{a}+O\left(r\varepsilon^2\right)~,
\ee
where $R=\sqrt{(X^1)^2+(X^2)^2+(X^3)^2}$.  Thus, we see that the potentials are the same as before, to Newtonian order, except that they now are in terms of comoving distance $X^i$ and conformal time $\eta$. We also now use Cartesian coordinates for simplicity.  

Our goal is to transform to the synchronous gauge, with new coordinates $\tilde{x}^{\mu}=(\tau,\tilde{x}^i)$, where the line element has the form
\be
ds^2=a^2(\tau)\tilde{g}_{\mu\nu}d\tilde{x}^{\mu}\tilde{x}^{\nu}=a^2(\tau)\left[-d\tau^2+\tilde{g}_{ij}d\tilde{x}^i\tilde{x}^j\right].
\ee
In this gauge, $\tilde{g}_{\tau\tau}=-1$ and $\tilde{g}_{\tau i}=\tilde{g}_{i\tau}=0$.  We make the following ansatz for the new coordinates:
\be
\eta=\tau+f_{(1)}\left(\tau,\tilde{x}\right)+f_{(2)}\left(\tau,\tilde{x}\right)+O\left(\tau_0\varepsilon^4\right)
\label{tau}
\ee
and
\be
X^i=\tilde{x}^i+h_{(1)}^i\left(\tau,\tilde{x}\right)+h_{(2)}^i\left(\tau,\tilde{x}\right)+O\left(\tilde{x}^i\varepsilon^2\right)~,
\label{tildex}
\ee
where $h_{(1)}^i\sim\delta\tilde{x}^i$, $h_{(2)}^i\sim\delta^2\tilde{x}^i$, $f_{(1)}\sim\delta\tau_0\varepsilon^2$, $f_{(2)}\sim\delta^2\tau_0\varepsilon^2$, and $\tau_0\sim\eta_0$ is the time today.  We are also assuming that we have the growing mode only, for which we have the power law scalings $\Phi_{(1)}\propto \tau^0$ and $\Phi_{(2)}\propto\tau^2$.

In order to find the new metric, we will need the relations
\be
a^2(\eta)=a^2(\tau)\left[1+\frac{4}{\tau}f_{(1)}+\frac{4}{\tau}f_{(2)}+O\left(\varepsilon^4\right)\right]
\ee
and
\be
\Phi_{(1)}(\eta,X) + \Phi_{(2)}(\eta,X) = \Phi_{(1)}(\tau,{\tilde x}) + \Phi_{(2)}(\tau,{\tilde x}) + \Phi_{(1),i} h^i_{(1)} + O(\varepsilon^4) + O(\delta^3)~.
\ee
Using these and the coordinate transformations (\ref{tau}) and (\ref{tildex}), we find
\be
\tilde{g}_{\tau\tau}=-\left(1+\frac{4}{\tau}f_{(1)}+\frac{4}{\tau}f_{(2)}+2\Phi_{(1)}+2\Phi_{(1),i}h_{(1)}^i+2\Phi_{(2)}+2\dot{f}_{(1)}+2\dot{f}_{(2)}\right)+{\dot h}_{(1)}^i {\dot h}_{(1)}^i=-1~,
\ee
implying
\be
\frac{2}{\tau}f_{(1)}+\Phi_{(1)}+\dot{f}_{(1)}=0
\label{diff1}
\ee
and
\be
\frac{4}{\tau}f_{(2)}+2\Phi_{(1),i}h_{(1)}^i+2\Phi_{(2)}+2\dot{f}_{(2)}-{\dot h}_{(1)}^i {\dot h}_{(1)}^i=0~.
\label{diff2}
\ee
Similarly, the time-space component of the new metric is
\be
\tilde{g}_{\tau i}=-f_{(1),i}-f_{(2),i}+\dot{h}_{(1)}^i+\dot{h}_{(2)}^i+h^j_{(1),i}\dot{h}_{(1)}^j+O\left(\varepsilon^3\right)=0
\ee
and this implies
\be
-f_{(1),i}+\dot{h}_{(1)}^i=0
\label{diff3}
\ee
and
\be
-f_{(2),i}+\dot{h}_{(2)}^i+h^j_{(1),i}\dot{h}_{(1)}^j=0~.
\label{diff4}
\ee
Equations (\ref{diff1}), (\ref{diff2}), (\ref{diff3}) and (\ref{diff4}) are solved by
\be
f_{(1)}=-\frac{\tau}{3}\Phi_{(1)}+\frac{A}{\tau^2}~,
\ee
\be
f_{(2)}=-\frac{\tau}{5}\Phi_{(2)}+\frac{2\tau^3}{45}\left(\nabla\Phi_{(1)}\right)^2+\frac{B}{\tau^2}- \frac{\tau}{6}  h_0^i \Phi_{(1),i}~,
\ee
\be
h_{(1)}^i=-\frac{\tau^2}{6}\Phi_{(1),i}+h_0^i\left(\tilde{x}\right)~,
\ee
and
\be
h_{(2)}^i=-\frac{\tau^2}{20}\Phi_{(2),i}+\frac{\tau^4}{120}\Phi_{(1),ij}\Phi_{(1),j}- \frac{\tau^2}{12}  \Phi_{(1),ji} h_0^j + \frac{\tau^2}{12} 
\Phi_{(1),j} h_{0,i}^j + {\tilde h}_0^i({\tilde x})~,
\ee
where the arbitrary constants $A$ and $B$ and functions $h_0^i(\tilde{x})$ and $\tilde{h}_0^i(\tilde{x})$ represent residual gauge freedoms associated with synchronous coordinates.  Setting $A$ and $B$ to zero will give us comoving coordinates.  We can imagine comoving coordinates to be fixed on some spacelike hypersurface from which the worldlines of freely falling particles emanate.  If we set all of the clocks carried by these particles to the same time on this spacelike hypersurface, then $A=B=0$.  The residual functions $h_0^i$ and $\tilde{h}_0^i$ correspond to simply changing the coordinates on the spacelike hypersurface from which worldlines emanate, and we will set $h_0^i=\tilde{h}_0^i=0$.  Using this solution for the appropriate coordinate transformation, we find the spatial part of the new metric to be
\ba
\tilde{g}_{ij}&=&\delta_{ij} \left[ 1 + \frac{4}{\tau} f_{(1)} + \frac{4}{\tau} f_{(2)} - 2 \Phi_{(1)} - 2 \Phi_{(2)} - 2 \Phi_{(1),k}h_{(1)}^k\right] - f_{(1),i} f_{(1),j}+ h_{(1)i,j} + h_{(1)j,i} \nonumber\\& &
+h_{(2)i,j} + h_{(2)j,i} + h_{(1)k,i} h_{(1)k,j}+ \left[ \frac{4}{\tau} f_{(1)} - 2 \Phi_{(1)} \right] \left[ h_{(1)i,j} + h_{(1)j,i}\right] + O(\varepsilon^4) + O(\delta^3)\nonumber\\
&=&\delta_{ij}-\frac{\tau^2}{3}\Phi_{(1),ij}-\frac{\tau^2}{10}\Phi_{(2),ij}+\frac{\tau^4}{60}\Phi_{(1),ijk}\Phi_{(1),k}+\frac{2\tau^4}{45}\Phi_{(1),ik}\Phi_{(1),jk}+O\left(\varepsilon^2\right)+O(\delta^3)~.
\ea

\end{document}